\documentclass[]{interact}
\newif\ifsupplement
\supplementfalse  

\usepackage{epstopdf}
\usepackage[caption=false]{subfig}
\usepackage[natbibapa,nodoi]{apacite}
\usepackage{amssymb} 
\usepackage{mathtools} 
\usepackage[colorinlistoftodos, textwidth = 20mm]{todonotes} 
\usepackage[ruled, vlined]{algorithm2e} 
\usepackage{multirow} 
\usepackage{siunitx} 
\usepackage{placeins} 
\usepackage[hidelinks]{hyperref}

\theoremstyle{plain}
\newtheorem{theorem}{Theorem}[section]

\newtheorem{assumption}[theorem]{Assumption}

\theoremstyle{definition}

\theoremstyle{remark}

\newcommand{\bbm}{\begin{bmatrix}}
\newcommand{\ebm}{\end{bmatrix}}
\newcommand{\wt}[1]{\widetilde{#1}}

\begin{document}

\articletype{ARTICLE}

\title{The Correlation Thresholding Algorithm for Exploratory Factor Analysis}


\author{
\name{D.~S. Kim, A.~Lu, and Qing~Zhou\thanks{CONTACT D.~S. Kim and Qing~Zhou; Email: daleskim@stat.ucla.edu, zhou@stat.ucla.edu. To appear in \emph{Structural Equation Modeling: A Multidisciplinary Journal}. This version reflects the originally submitted manuscript with minor typographical corrections and formatting changes.}}
\affil{Department of Statistics \& Data Science \\ University of California, Los Angeles}
}

\maketitle 

\begin{abstract}
Exploratory factor analysis is often used in the social sciences to estimate potential measurement models.
To do this, several important issues need to be addressed: (1) determining the number of factors, (2) learning constraints in the factor loadings, and (3) selecting a solution amongst rotationally equivalent choices.
Traditionally, these issues are treated separately.
This work examines the Correlation Thresholding (CT) algorithm, which uses a graph-theoretic perspective to solve all three simultaneously, from a unified framework.
Despite this advantage, it relies on several assumptions that may not hold in practice.
We discuss the implications of these assumptions and assess the sensitivity of the CT algorithm to them for practical use in exploratory factor analysis.
This is examined over a series of simulation studies, as well as a real data example.
The CT algorithm shows reasonable robustness against violating these assumptions and very competitive performance in comparison to other methods.

\end{abstract}


\section{Introduction}

Exploratory factor analysis (EFA) is a common procedure for estimating a factor analysis model in the absence of a priori theory.
To perform EFA, three issues must be addressed: (1) determine the number of factors, (2) learn the constraints in the factor loadings, (3) determine a solution amongst rotationally equivalent choices.
Typically, these three problems are addressed separately, leading to a large set of possible combinations, all with different solutions.
For example, the number of factors can be chosen via a variety of eigenvalue methods \citep{Cattell1966, Guttman1954, Glorfeld1995, Horn1965, Kaiser1960, Raiche2013} or a model selection approach \citep{Preacher2013}, factor loadings are often set to zero if they are below an ad-hoc threshold \citep{Ford1986, Howard2016}, and there are numerous criteria for rotational identifiability \citep[for a review see][]{Browne2001}.
Moreover, none of these methods are without controversy, and a great deal of literature has been devoted to criticisms on both empirical and theoretical grounds \citep{Browne1968,Ford1986, Zwick1986, Velicer1990, Howard2016, Auerswald2019}.

More recently, penalized EFA methods have been developed to partially address these issues.
The most familiar are the LASSO \citep{Tibshirani1996} and MCP \citep{Zhang2010} penalties used in a penalized likelihood method developed by \cite{Hirose2014b}.
Instead of rotating factor coefficients, penalized EFA can achieve sparse solutions directly in the estimation of the factor loadings.
This requires the use of tuning parameters, followed by model selection with the Bayesian Information Criterion (BIC) or cross-validation \citep[CV;][]{Scharf2019}.
While these procedures are more cohesive than traditional EFA, they require a grid search over a large set of tuning parameters which is computationally intense.
Furthermore, the number of factors is still required as an input, and theoretical guarantees for rotational identifiability and structural estimation consistency have yet to be established for many penalty functions.

To solve all these problems from a unified framework, recently \cite{Kim2023} proposed a 
graph-theoretic solution, using a correlation thresholding (CT) algorithm.
The main idea centers around the heuristic that correlations between variables that share a common factor (within-factor correlation) are usually stronger than correlations between variables that do not (between-factor correlation).
If this assumption holds, a correlation graph that removes (or thresholds) the between-factor correlations will correspond to the constraints of the factor loadings, thereby recovering the factor analysis model structure.
Factor analysis models that imply such correlation matrices are termed \textit{thresholdable}.
Further theoretical results are obtained if there is a unique indicator variable per latent factor, which is termed the \textit{unique child condition}.
If the factor analysis model is thresholdable and the unique child condition holds, then model structures output by the CT algorithm are statistically consistent, under both low- and high-dimensional regimes.
The unique child condition also guarantees the model structure is rotationally unique.

It is important to note that these two assumptions are sufficient but not necessary conditions for the consistency guarantees of the CT algorithm. It is possible that the CT algorithm can give accurate structure learning for a factor model even if one or both assumptions do not hold.
As such, to determine the practical use of the CT algorithm as an EFA tool, in this article, we conduct a systematic study of the sensitivity and robustness of this procedure to its assumptions, including comparisons with other structure learning methods.

To support this investigation, we now provide a formal specification of the factor analysis model.
Let $X = (X_1, \dots, X_p)$ be a function of unobserved latent factors $L = (L_1, \dots, L_d)$.
This is typically specified as a set of linear structural equations:
\begin{equation} \label{model}
X = \Lambda L + \epsilon,
\end{equation}
where $L \sim \mathcal{N}_d (0, \Phi)$, $\epsilon$ is a vector of uncorrelated error terms $(\epsilon_1, \dots, \epsilon_p) \sim \mathcal{N}_p (0, \Omega)$, and $\Lambda = (\lambda_{ij}) \in \mathbb{R}^{p \times d}$ is a matrix of factor loadings.
Without loss of generality, we omit an additive mean vector $\mu$ from the model.
Since factor analysis is typically used as a dimension simplification technique, we assume that $d < p$.

The factor analysis model in Equation~\ref{model} implies a covariance structure $\Sigma$ for $X$ as follows:
\begin{equation} \label{eq:varexp}
\Sigma(\theta) \coloneqq \Lambda \Phi \Lambda^T + \Omega,
\end{equation}
letting $\theta = \{\Lambda, \Phi, \Omega\}$.
It will also be relevant to work with the correlation matrix of $X$.
Let $D_{\sigma}=\text{diag}(\Sigma)^{1/2}$, i.e. a diagonal matrix with entries $\Sigma_{ii}^{1/2}$.
Then we define a unit variance scaled $X$ as $\wt{X}$ in the following manner:
\begin{equation}
\wt{X} \coloneqq D^{-1}_{\sigma} X = D^{-1}_{\sigma}(\Lambda L + \epsilon) = \wt{\Lambda} L + \wt{\epsilon},
\end{equation}
where $\wt\Lambda=D^{-1}_{\sigma}\Lambda$ and $\wt{\epsilon}=D^{-1}_{\sigma} \epsilon$.
Similarly, it follows that a correlation matrix $\wt{\Sigma}$ can be expressed as:
\begin{equation}\label{eq:Sigmaexp}
\wt{\Sigma}(\theta) \coloneqq D^{-1}_{\sigma} \Sigma D^{-1}_{\sigma} = \widetilde{\Lambda} \Phi \widetilde{\Lambda}^T + \widetilde{\Omega},
\end{equation}
where $\wt{\Omega} = D^{-1}_{\sigma} \Omega D^{-1}_{\sigma}$, and the elements of $\wt{\Sigma}(\theta)$ may be referred to as $\rho_{ij}$.
Finally, notice that the structure of a factor analysis model is entailed by the number of factors $d$ and the support of $\Lambda$ (or equivalently its constraints), denoted $\mathcal{A}(\Lambda)$.
Therefore we will formally define the \textit{structure} of a factor analysis model as the pair $(d, \mathcal{A}(\Lambda))$.
This formalization establishes the foundation of the problem, which is the recovery of the structure $(d, \mathcal{A}(\Lambda))$ from observed data.

The remainder of this article is organized as follows.
In Section 2, we introduce the CT algorithm and review its theoretical properties.
In Section 3, we critically discuss the algorithm's assumptions and their implications.
Section 4 presents simulation studies designed to evaluate sensitivity to these assumptions in both low- and high-dimensional settings.
Section 5 illustrates the CT algorithm using real data, and Section 6 concludes with a discussion and directions for future research.

\section{The Correlation Thresholding Algorithm} \label{sec:ct_algorithm}

\subsection{Overview} \label{sec:ct_overview}

\begin{figure}[tb]
  \centering
  \includegraphics[width=0.4\textwidth, keepaspectratio]{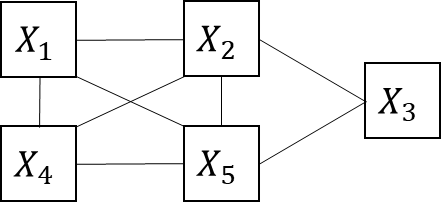}
  \caption{An example graph for illustrating cliques.
  The vertex set $\{X_1, X_2, X_4\}$ is a clique, but not a maximal clique, since $X_5$ can be added to the set and it would remain a clique.
  The set $\{X_1, X_2, X_4, X_5\}$ is a maximal clique, since there is no vertex that can be added to it and have the set remain a clique.
  }
  \label{fig:graph_example}
\end{figure}

To begin, we review several terms and definitions from graph theory.
We define a graph $\mathcal{G}$ as an ordered pair $(V, E)$, explicitly denoted as $\mathcal{G}(V, E)$, where $V$ is a set of vertices and $E \subseteq V \times V$ is a set of edges.
For convenience, we will use $V = X$ to mean that the vertex set $V$ represents the random vector $X$.
We also restrict our attention to \textit{undirected} graphs, where we only consider edges $(i, j) \in E$ such that $i < j$.
A \textit{clique} of $\mathcal{G}(V, E)$ is a subset of vertices $C \subseteq V$ such that all pairs of distinct vertices in $C$ are connected by an edge.
A \textit{maximal clique} is a clique that cannot be extended by including more vertices from $V$ (i.e., not a subset of any other clique).
An example that illustrates the difference between a clique and maximal clique is provided in Figure~\ref{fig:graph_example}.
Finally, define the \textit{parent set} of an observed variable $X_j$ as $pa(X_j) \coloneqq \{k : \lambda_{jk} \neq 0\}$ and likewise the \textit{child set} of a latent variable $L_k$ as $ch(L_k) \coloneqq \{j : \lambda_{jk} \neq 0\}$.

\begin{figure}
\centering
\subfloat[]{%
\resizebox*{.3\textwidth}{!}{\includegraphics{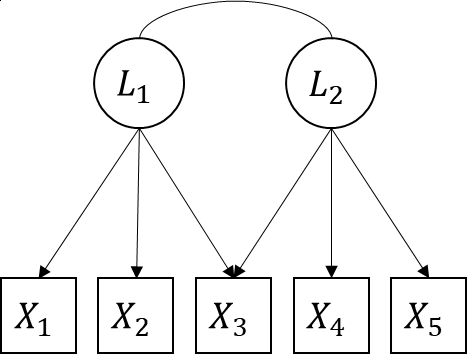}}}\hspace{5pt}
\subfloat[]{%
\resizebox*{.3\textwidth}{!}{\includegraphics{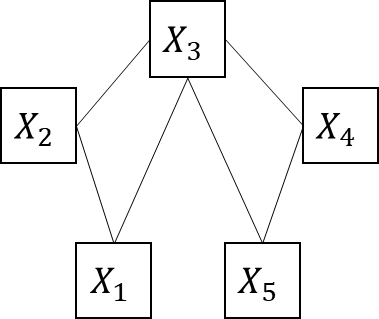}}}
  \caption{
  (a) The factor analysis model described in Equation~\ref{e_example}; (b) The graph $\mathcal{G}(X, E_0)$.
  } \label{fig:graph_demo}
\end{figure}

The CT algorithm is based off the heuristic that correlations amongst variables that have a latent factor in common (within-factor correlations) tend to be stronger than correlations amongst variables that do not share a common factor (between-factor correlations).
Under certain assumptions (discussed below), \cite{Kim2023} showed that the factor analysis structure can be recovered by analyzing graphs corresponding to the thresholded correlation matrix.
This can be demonstrated with a simple example.
Consider the following parameters:
\begin{equation} \label{e_example}
\wt{\Lambda} = \begin{bmatrix}
\wt{\lambda}_{11} & \\
\wt{\lambda}_{21} & \\
\wt{\lambda}_{31} & \wt{\lambda}_{32} \\
 & \wt{\lambda}_{42} \\
 & \wt{\lambda}_{52} \\
\end{bmatrix},\,
\Phi = \begin{bmatrix}
1 & \phi_{12} \\
\phi_{12} & 1\\
\end{bmatrix},\,
\wt{\Omega} = \begin{bmatrix}
\wt{\omega}_1&&&&\\
&\wt{\omega}_2&&&\\
&&\wt{\omega}_3&&\\
&&&\wt{\omega}_4&\\
&&&&\wt{\omega}_5
\end{bmatrix}.
\end{equation}
This model is illustrated in Figure~\ref{fig:graph_demo}a.
To begin, we inspect the entries of $\wt{\Sigma}(\theta) = \wt{\Lambda} \Phi \wt{\Lambda}^T$, whose non-redundant elements are
$$\small\label{e_example1}
\bbm
\wt{\lambda}^2_{11} + \wt{\omega}^2_1 & & & & \\[3pt] 
\wt{\lambda}_{11} \wt{\lambda}_{21} & \wt{\lambda}^2_{21} + \wt{\omega}^2_2 & & & \\[3pt] 
\wt{\lambda}_{11} \wt{\lambda}_{31} + \wt{\lambda}_{11} \wt{\lambda}_{32} \phi_{12} \text{   } & \text{   } \wt{\lambda}_{21} \wt{\lambda}_{31} + \wt{\lambda}_{21} \wt{\lambda}_{32} \phi_{12} \text{   } & \text{   } \wt{\lambda}^2_{31} + \wt{\lambda}^2_{32} + \wt{\omega}^2_3 \text{   } & & \\[3pt] 
\bm{\wt{\lambda}_{11} \wt{\lambda}_{42} \phi_{12}} & \bm{\wt{\lambda}_{21} \wt{\lambda}_{42} \phi_{12}} & \text{   } \wt{\lambda}_{31} \wt{\lambda}_{42} \phi_{12}+ \wt{\lambda}_{32} \wt{\lambda}_{42} \text{   }  & \text{   } \wt{\lambda}^2_{42} + \wt{\omega}^2_4 \text{   } & \\[3pt] 
\bm{\wt{\lambda}_{11} \wt{\lambda}_{52} \phi_{12}} & \bm{\wt{\lambda}_{21} \wt{\lambda}_{45} \phi_{12}} & \text{   }  \wt{\lambda}_{31} \wt{\lambda}_{52} \phi_{12}+ \wt{\lambda}_{32} \wt{\lambda}_{52} \text{   } & \wt{\lambda}_{42} \wt{\lambda}_{52} & \text{   } \wt{\lambda}^2_{52} + \wt{\omega}^2_5 
\ebm.$$
The key observation is that the between-factor correlations (bolded) all must have a $\phi_{ij}$ shrinkage factor.
Conversely, the within-factor correlations will all have at least one term without the $\phi_{ij}$ shrinkage factor.
Suppose that we use a graph to only encode the within-factor correlations.
In graph terminology, this leads to the edge set of
\begin{equation} \label{eq:pop_graph}
E_0 \coloneqq \{(i, j) : pa(X_i) \cap pa(X_j) \neq \emptyset\}.
\end{equation}
That is, a pair of variables must share a common factor if the intersection of their parent sets are non-empty.
This graph is displayed in Figure~\ref{fig:graph_demo}b.
Note that we can conversely describe the pairs of variables that do not share a common factor with the complement of $E_0$ as
\begin{equation} \label{eq:pop_graph_c}
E_0^c = \{(i, j) : pa(X_i) \cap pa(X_j) = \emptyset\}.
\end{equation}

The graph $\mathcal{G}(X, E_0)$ possesses an interesting correspondence with the underlying factor analysis structure.
First, the number of latent variables ($d=2$) equals the number of maximal cliques in $\mathcal{G}(X, E_0)$.
Second, these maximal cliques correspond to the children sets of the latent variables in the factor analysis model.
In Figure~\ref{fig:graph_demo}b, the maximal cliques $\{1, 2, 3\}$ and $\{3, 4, 5\}$ correspond to the respective children sets of $L_1$ and $L_2$ as can be seen in Figure~\ref{fig:graph_demo}a.
The correlation thresholding algorithm leverages this correspondence by using these graphs to gain insight into the structure of the factor analysis model.

If indeed the between-factor correlations have a lower magnitude than the within-factor correlations, then there will exist a threshold $\tau_0$ such that the correlations corresponding to $E_0$ and $E_0^c$ can be separated.
If a set of parameters $\theta$ admits such a separation, we call $\theta$ \textit{thresholdable}.
Formally, this condition is described as follows.
\begin{assumption}[Thresholdable] \label{as:thresholdable}
Let $\theta$ be a set of parameters for a factor analysis model and $E_0$ be the set of index pairs defined in Equation~\ref{eq:pop_graph}.
If there exists a $\tau_0$ such that
\begin{equation} \label{eq:thresholdable}
\max\{ \lvert \rho_{kl} \rvert : (k, l) \in E_0^c \} < \tau_0 < \min\{ \lvert \rho_{ij} \rvert : (i, j) \in E_0 \},
\end{equation}
then $\theta$ is called \emph{thresholdable}.
\end{assumption}

That is, $\theta$ is thresholdable if the smallest within-factor correlation is greater than the largest between-factor correlation, in terms of magnitude.
Therefore, we can consider a \textit{thresholded correlation graph} $\mathcal{G}(X, E(\tau))$, where the edge set is determined by thresholding the values of $\lvert \rho_{ij} \rvert$ at $\tau\geq 0$, i.e.,
\begin{equation} \label{eq:thresholded_graph}
E(\tau) \coloneqq \{(i, j) :  |\rho_{ij}| > \tau\},
\end{equation}
and hence given $\tau_0$ we would have $E(\tau_0) = E_0$.
Subsequently, this leads to a natural sample estimate of $E_0$
\begin{equation} \label{eq:E_hat}
\hat{E}(\tau) \coloneqq \{(i, j) : \lvert r_{ij} \rvert > \tau\},
\end{equation}
where $r_{ij}$ denotes the sample correlation.

Therefore, given a thresholdable $\theta$, this distills the problem of learning a factor analysis structure into a search for a suitable $\tau_0$.
This can be accomplished by checking a set of candidate thresholds $\tau_k \in (0, 1)$ and analyzing each of the graphs $\mathcal{G}(X, \hat{E}(\tau_k))$.
In turn, these graphs give us a set of candidate factor analysis structures, from which we can estimate a set of models to choose from.
A final model can then be selected by using a model selection procedure (e.g., BIC).

\subsection{The Algorithm}

The intuition and logic from corresponding thresholded correlation graphs and factor analysis structures can be built into an formal algorithm as described in Algorithm~\ref{alg:ct}.
The algorithm begins with a sample correlation matrix $R$ and a set of thresholds $\tau = \{\tau_1, \dots, \tau_m\}$ to test.
Then for each threshold $\tau_k$, we analyze $\mathcal{G}(X, \hat{E}(\tau_k))$ for independent maximal cliques.
The number of independent maximal cliques is then taken as our number of latent factors and the members of these cliques are correspondingly the children of these factors.
This yields a set of $m$ structures that can be estimated (e.g., via MLE) from which model selection can be conducted to yield a final model (e.g., BIC).

\begin{algorithm}[!htb]
\SetAlgoLined
\LinesNumbered
\SetKwInOut{Input}{input}
\SetKwInOut{Output}{output}
\Input{The sample correlation matrix $R$ and a set of thresholds $\tau$.}
\Output{Parameter estimates $\hat{\theta}$.}
 \For{$k \in [m]$}{
   Calculate $\mathcal{G}(X, \hat{E}(\tau_k))$ and extract the set of independent maximal cliques: $\mathcal{C}_k = \{C_1, \dots C_{\lvert \mathcal{C}_k \rvert}\}$\;\nllabel{ct_Ehat}
   Set $\hat{d}_k = \lvert \mathcal{C}_k \rvert$\;\nllabel{step:learn_d}
   Initialize $\hat{\mathcal{A}}_k = \emptyset$\;\nllabel{step:initialize_A}
   \For{$(i, j) \in [p] \times [\hat{d}_k]$\nllabel{forlambda}}{
   \lIf{$i \in C_j$}{add $(i, j)$ to $\hat{\mathcal{A}}_k$\nllabel{step:learn_a}}
   }\nllabel{step:end_support_learning}
 Estimate $\hat{\theta}_k$ given $(\hat{d}_k, \hat{\mathcal{A}}_k)$, i.e. subject to $\lambda_{ij}=0$ for all $(i, j) \notin \hat{\mathcal{A}}_k$\;\nllabel{ct_estimate}
 }
 Select one of the $m$ estimates from $\{ \hat{\theta}_k : k \in [m] \}$ via a model selection procedure.\nllabel{ct_select}
 \caption{The Correlation Thresholding Algorithm}
\label{alg:ct}
\end{algorithm}

We first note the use of \textit{independent} maximal cliques in Step~\ref{ct_Ehat} of the algorithm.
An independent maximal clique, is a maximal clique that contains a vertex not shared by any other maximal clique.
This simply means that independent maximal cliques have at least one vertex that is unique to them.
This is a computational necessity, as discovering all maximal cliques has an exponential computational complexity (with respect to $p$) while a search for independent maximal cliques can be done in polynomial time \citep{Eppstein2010, Kim2023}.
This has some implications on the identification of factor analysis structures which are discussed in the next section.

Once the set of independent maximal cliques are extracted, Step~\ref{step:learn_d} through~\ref{step:end_support_learning} convert the graph to a factor analysis structure.
The number of independent maximal cliques is set as the estimate of the number of latent factors, and the members of each clique determine if a loading should be free or constrained in $\hat{\Lambda}$.
Parameters are then estimated in Step~\ref{ct_estimate} based on this estimated structure.

\begin{figure}[tb]
  \centering
  \includegraphics[width=\textwidth, keepaspectratio]{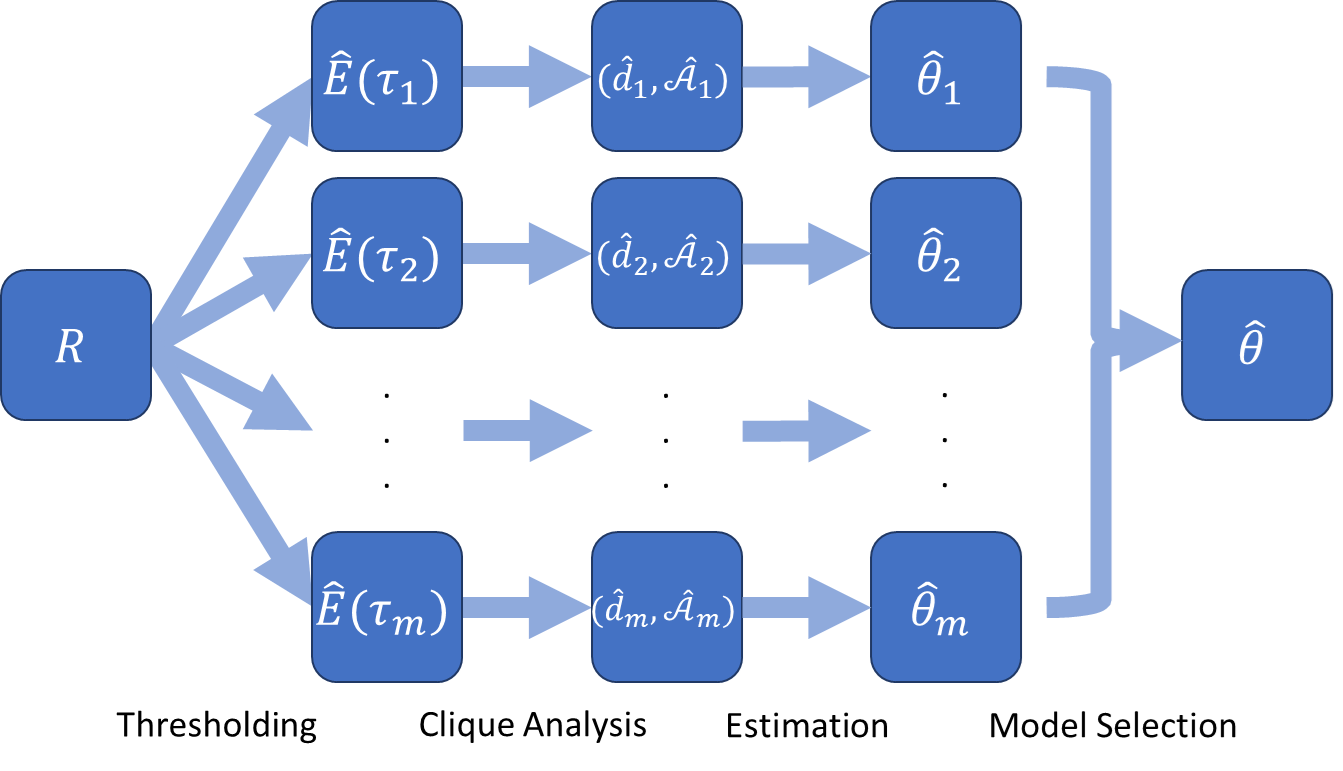}
  \caption{A high-level diagram of the CT algorithm. A set of candidate thresholds $\tau_k$ each suggest a candidate factor analysis structure through clique analysis. Each of these structures is then estimated and compared through model selection.}
  \label{fig:alg}
\end{figure}

We provide a high-level summary of the algorithm in Figure~\ref{fig:alg}.
The main assertion of the algorithm is that under a good choice of input thresholds, the correct structure will be recovered by one of the graphs $\mathcal{G}(X, \hat{E}(\tau_k))$.
Then, given the correct model is among the final set of candidate models, a consistent model selection criterion will be able to recover it.

\cite{Kim2023} showed that solutions provided by the CT Algorithm enjoy several desirable statistical properties.
First, consistency of both the structure estimate and parameter estimates will hold so long as a consistent parameter estimation method is used in Step~\ref{ct_estimate}, and a consistent model selection procedure is used in Step~\ref{ct_select}.
This is easily satisfied by using MLE and BIC model selection.
More interestingly, these consistency results hold in high-dimensional settings as well, where $p$ may grow with $n$ and even $p\gg n$.
Moreover, all solutions chosen by the CT algorithm are also rotationally unique.
These results are established under two key assumptions: the thresholdability assumption and the unique child condition.
We discuss these assumptions in further detail in the next section.

\section{Discussion of Assumptions} \label{sec:assumptions}

We discuss two key assumptions for consistent recovery of latent factor model structures by the CT algorithm.
Since these assumptions may not hold in all practical applications, we examine the sensitivity and robustness of the CT algorithm when these assumptions are violated to different degrees using systematical simulation studies in Section~\ref{sec:simulations}.

\subsection{Thresholdability}

The thresholdability assumption (Assumption~\ref{as:thresholdable}) formalizes the heuristic that within-factor correlations are stronger than between-factor correlations.
This tends to hold in practice, as several aspects of thresholdability are designed into factor analysis items.
First, it is more optimal for a measurement model to have higher within-factor correlations due to higher reliability \citep{Graham2006}.
Second, it is desirable that factors are distinguishable from each other, or not highly correlated \citep{Whitely1983}, or else a structure with less factors may be better suited.
Also note that if the latent factors are orthogonal, thresholdability will trivially hold.

Additionally, \cite{Kim2023} showed that exceptions to thresholdability are much less likely to occur under so-called ``independent cluster structures'' \citep{Harris1964}, which is another common factor analysis design.
These are factor analysis structures where observed variables cluster into mutually exclusive sets per latent variable (i.e., no cross-loadings), and have also been called ``perfect simple structures'' \citep{Jennrich2006}.
Such structures are ideal for their easily interpreted measurement structure, and is an important design in educational and psychological test construction \citep{Hattie1985, Anderson1988}.
Despite these design elements, it has been pointed out that exceptions are not completely ruled out \citep{Bollen1991}.
The degree to which the CT algorithm is sensitive to violations of thresholdability is what we will examine in the simulation studies.

\subsection{Unique Child Condition}

The use of independent maximal cliques in Algorithm~\ref{alg:ct} to recover the support of $\Lambda$ depends on a so-called ``unique child condition'' \citep{Kim2023}.
This is formally defined as follows.
\begin{assumption}[Unique Child Condition] \label{as:ucc}
Let $U_k$ be the set of unique indicators of the latent variable $k$.
Then if all $U_k$ are non-empty:
\begin{equation} \label{eq:uc_def}
U_k \coloneqq \text{\emph{ch}}(L_k) - \bigcup_{j \neq k} \text{\emph{ch}}(L_j) \neq \emptyset,\quad\quad \forall\; k\in\{1, ..., d\},
\end{equation}
then we say that the \emph{unique child condition} holds.
\end{assumption}

In other words, it assumes that each latent factor has at least one observed variable that uniquely indicates it (i.e., at least one variable with no cross-loadings).
This can be thought of as a minimal version of an independent cluster structure, where the cluster structure only needs to hold for at least one observed variable per latent variable, and the rest of the structure can be freely saturated.
This is a much more relaxed assumption than the independent cluster structure, which is required by other structure learning algorithms \citep{Scheines1998, Jennrich2001, Jennrich2006, Silva2006}, or algorithms that require 3 to 4 unique children per latent variable \citep{Shimizu2009, Kummerfeld2016}.
If the unique child conditions holds, then there is a one-to-one correspondence between the independent maximal cliques and the child sets of each latent factors.
This means that each latent factor is uniquely identified by each independent maximal clique.

\begin{figure}
\centering
\subfloat[]{%
\resizebox*{.3\textwidth}{!}{\includegraphics{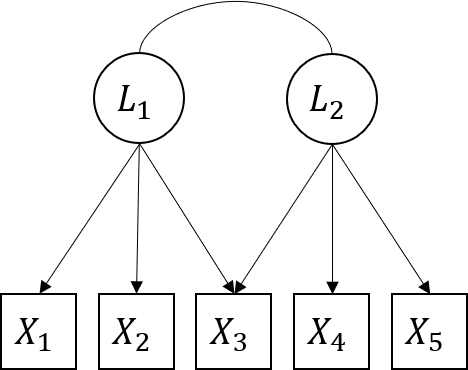}}}\hspace{5pt}
\subfloat[]{%
\resizebox*{.3\textwidth}{!}{\includegraphics{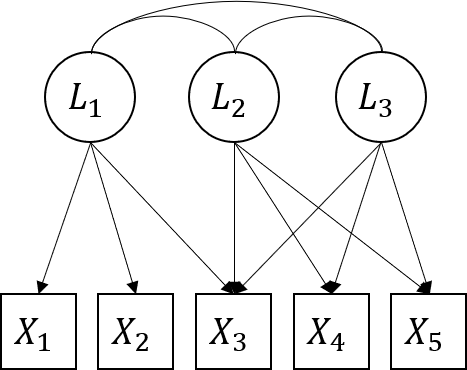}}}\hspace{5pt}
\subfloat[]{%
\resizebox*{.3\textwidth}{!}{\includegraphics{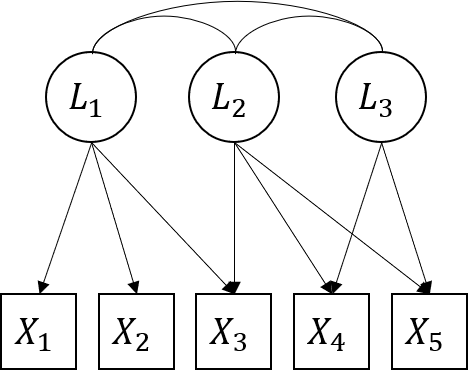}}}
\caption{Three structures that yield the same thresholded correlation graph.} \label{fig:ucc_example}
\end{figure}

However, the unique child condition does impose some restrictions on the types of factor analysis model structures.
We illustrate several examples in Figure~\ref{fig:ucc_example}.
Assuming each of the displayed models is thresholdable with $\tau_0$, all these structures will yield the same thresholded correlation graph with edge set $E(\tau_0)$.
Specifically, the independent maximal cliques that are yielded by them are $\{1, 2, 3\}$ and $\{3, 4, 5\}$, despite all having different structures.
This can be seen by noting that some latent variables do not yield maximal cliques in $\mathcal{G}(X, E_0)$, or yield the same independent maximal clique as another latent variable.
For example, in Figure~\ref{fig:ucc_example}b, both $L_2$ and $L_3$ yield the clique $\{3, 4, 5\}$.
Thus, $L_2$ and $L_3$ cannot be distinguished from each other through independent maximal cliques alone.
Similarly, in Figure~\ref{fig:ucc_example}c, $L_3$ yields the clique $\{4, 5\}$, but it is not maximal since $L_2$ yields $\{3, 4, 5\}$.
In this case, $L_3$ will not be identified as a latent variable, since its clique is subsumed by the one yielded by $L_2$.
Therefore, the unique child condition is required for perfect recovery of a model structure by independent maximal cliques in a correlation thresholded graph.
In our simulation studies, we examine the effect to which structure recovery is hampered when this assumption is violated.

\section{Simulation Studies} \label{sec:simulations}

\subsection{Simulation Design and Comparison Metrics} \label{sec:metrics}

In the simulation studies, we compare the CT algorithm against three other methods: (1) EFA, (2) EFA-LASSO, and (3) EFA-MCP.
Note that these EFA methods all require $d$ as an input, thus we use the CT algorithm to give these EFA methods a set of $d$ to work with.
This was to make the comparison as fair as possible, rather than using ad-hoc choices.
More specifically, we ran the CT algorithm to Step~\ref{step:learn_d}, where $d$ is estimated from the number of independent maximal cliques ($\hat{d}_k$).
Thereafter, we replaced the constraint learning portion (Steps~\ref{step:initialize_A} through~\ref{step:end_support_learning}) of the algorithm with one of the EFA procedures.
Then the structure of the model was saved from the EFA methods and resumed the algorithm from Step~\ref{ct_estimate}, where the MLE was estimated and used for model selection.
For the thresholds $\tau_k$, we used all unique sample values $\lvert r_{ij} \rvert$.
This is feasible for low-dimensional settings as the number of thresholds will be no greater than $p(p - 1)/2$.
All data sets were generated from a zero-mean Gaussian distribution, with a covariance matrix $\Sigma$ parameterized by $\theta$.

The simulations were done in the \texttt{R} language \citep[4.3.1;][]{R2023}.
The \texttt{OpenMx} package \citep{Neale2016} was used to estimate the MLEs of all methods.
We used the \texttt{fanc} package \citep{Hirose2014a, Hirose2014b} to learn the structure of the LASSO and MCP variants of EFA.
The tuning parameters were left at the package defaults of 30 values for a single tuning parameter in LASSO and 270 combinations of two tuning parameters in MCP.
We omitted the rotation and constraint learning steps for the traditional EFA method, and used it as a saturated model baseline.
All code for the simulations are included in the supplementary materials, linked in Appendix~\ref{app:osf}.

We examined several outcomes to assess the performance of the methods in terms of three categories: (1) structural accuracy, (2) model fit, and (3) computational efficiency.
For structural accuracy, we collected Hamming distance (HD), the $F_1$ score of $\mathcal{A}(\hat{\Lambda})$ and the estimated number of latent factors.
The Hamming distance is a measure of discrete error, which is the number of substitutions to make two objects the same.
In the context of $\mathcal{A}(\hat{\Lambda})$, it is the number of errors made in estimating $\mathcal{A}(\Lambda)$.
Since Hamming distance may depend on the size of $\Lambda$, we also used a normalized version of accuracy in the $F_1$ score.
The $F_1$ score is the harmonic mean of precision and recall for classification problems, where we view $\mathcal{A}(\hat{\Lambda})$ as a classifier for the paths between $X$ and $L$.
It can be equivalently be expressed in terms of true positives (TP), false positives (FP), and false negatives (FN) as follows:
\begin{equation}
F_1 = \dfrac{2\text{TP}}{2\text{TP} + \text{FP} + \text{FN}}.
\end{equation}
Some additional computational adjustments are made for HD and $F_1$ to account for the fact that column order of $\hat{\Lambda}$ may not be the same as $\Lambda$.
These details are presented in Appendix~\ref{app:struc_acc}.
We also collected the estimated number of latent factors, $\hat{d}$.

For model fit, we collected the Tucker-Lewis Index (TLI; \citealp{Tucker1973}) and the root mean square error of approximation (RMSEA; \citealp{Steiger1980}).
We also generated a second dataset from the true model and calculated a test data log-likelihood with the estimated model.
To measure computational efficiency we simply counted the number of models each method estimated.
This was to avoid idiosyncratic differences between the software implementations of each method.
For the CT algorithm, this is simply the number of unique structures obtained by the sequence of $\tau_k$.
For EFA, this translates to the number of unique $d$ obtained by the sequence of $\tau_k$.
For EFA-LASSO and EFA-MCP, this is the number of tuning parameter combinations to search over (30 for LASSO, 270 for MCP), per unique $d$ in the sequence of $\tau_k$.

\subsection{Violations to Thresholdability} \label{sec:thresh_sim}

For our first simulation study we examined violations to the thresholdability assumption in the low-dimensional setting.
The number of latent variables took on values $d \in \{2, 3, 4, 5\}$ with the number of children per latent variable set to 5.
We set $n \in \{250, 500, 1000\}$ and generated 100 datasets per condition.
We will only examine the results for the $n = 1000$ case, since the $n = 250$ and $n = 500$ conditions are quite similar in their patterns of results.
The results for the other sample sizes are provided in the supplementary materials, linked in Appendix~\ref{app:osf}.

We violated the thresholdability assumption by modifying the degree to which the factors were correlated.
Each of the $\lambda_{ij}$ and $\phi_{jk}$ parameters were essentially drawn from a range of $[0.6, 0.8]$, whose details can be found in Appendix~\ref{app:data_gen}.
Then the relative magnitude of the factor correlations was multiplied by a scaling factor $\alpha \in \{0, 0.25, 0.5, 0.75, 1\}$.
This controlled the frequency at which the generated $\theta$ was thresholdable, where $\alpha = 0$ guaranteed thresholdability, and $\alpha = 1$ severely diminished it.
For the structure of $\Lambda$, we began with an independent cluster structure (one non-zero entry per row), then randomly added cross-loadings to half of the observed variables to better emulate real data settings.
To avoid confounding, the unique child condition was enforced by designating one observed variable per latent variable where cross-loadings were disallowed.

For each generated $\theta$, we tested thresholdability by applying the definition in Assumption~\ref{as:thresholdable}.
To examine a more continuous measure of thresholdability we also checked the proportion of within-factor correlations that were higher than the maximum between-factor correlation and between-factor correlations that were lower than the minimum within-factor correlation.
This essentially counts how many correlations can always be correctly sorted as within-factor or between-factor using a best case threshold.
We term this measure \textit{sortability} and provide a formal definition in Appendix~\ref{app:sortability}.

\begin{figure}[tb]
  \centering
  \includegraphics[width=\textwidth, keepaspectratio]{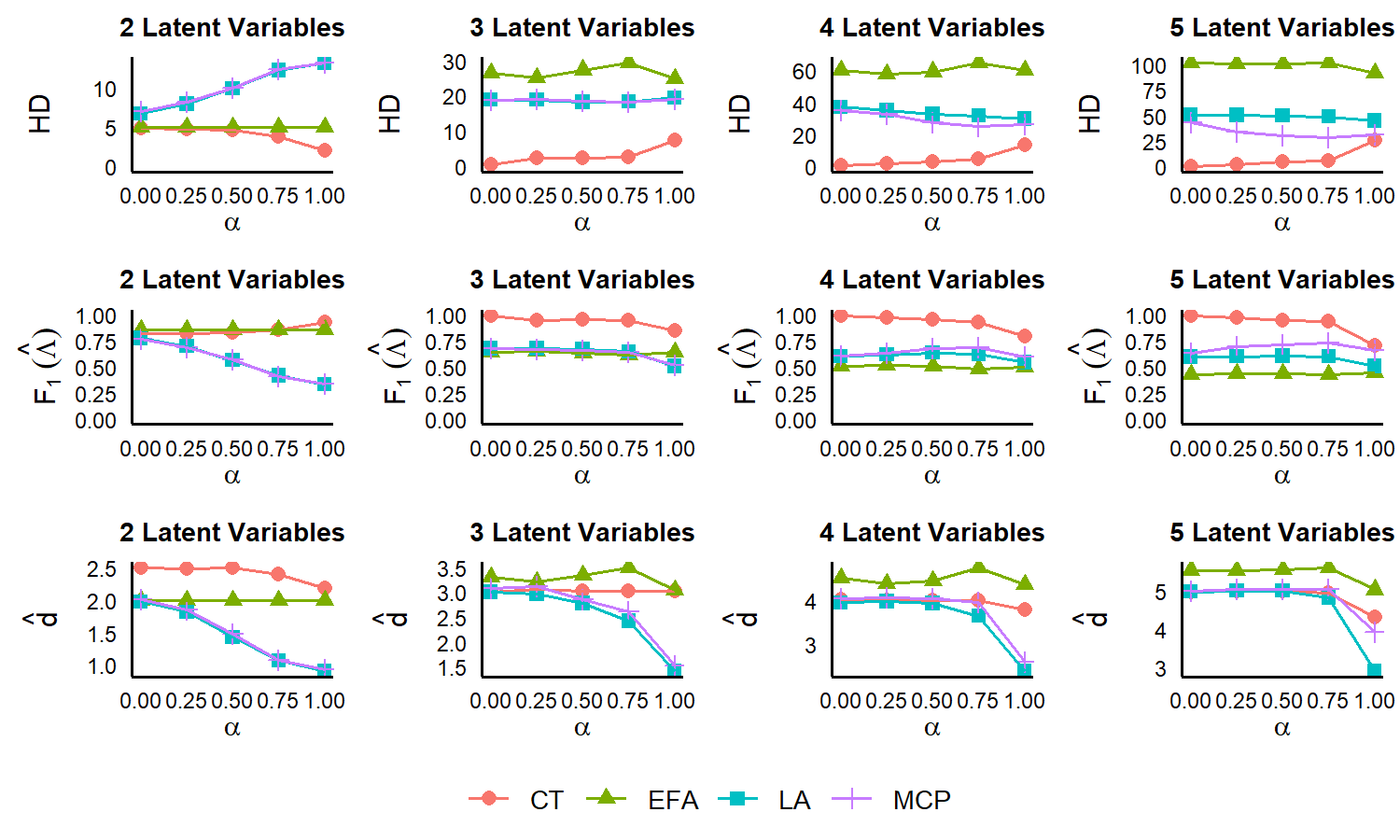}
  \caption{Average structural accuracy statistics for the low-dimensional simulation, as a function of $\alpha$.}
  \label{fig:alpha_str_1000}
\end{figure}

\begin{figure}[tb]
  \centering
  \includegraphics[width=\textwidth, keepaspectratio]{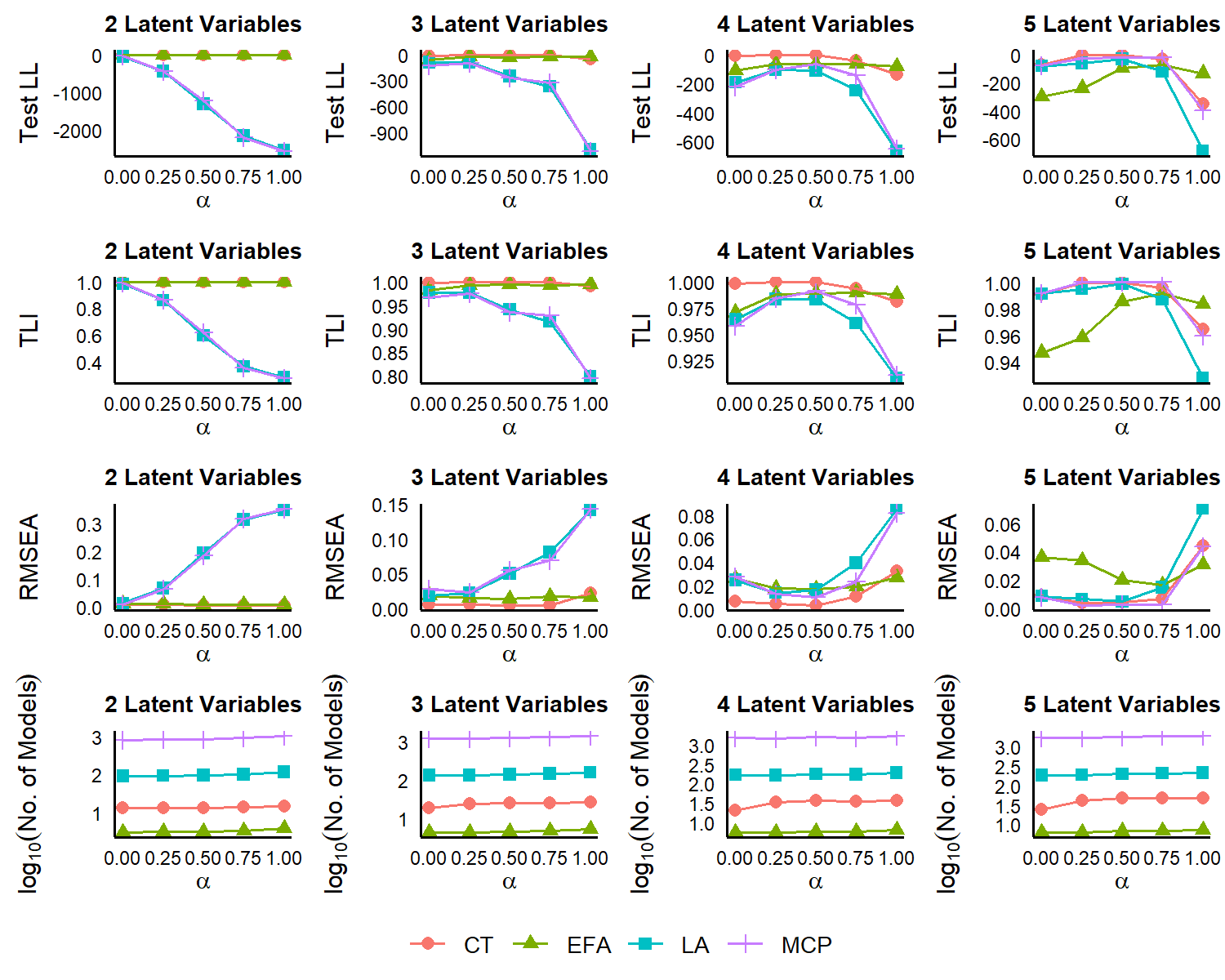}
  \caption{Average model fit and computational efficiency statistics for the low-dimensional simulation, as a function of $\alpha$.
  For the fit statistics in $d = 2$, we note that CT and EFA are largely overlapping, and LA and MCP are largely overlapping.}
  \label{fig:alpha_fit_1000}
\end{figure}

The results for varying $\alpha$ are displayed in Figure~\ref{fig:alpha_str_1000} for the structural accuracy outcomes and in Figure~\ref{fig:alpha_fit_1000} for the statistical fit and computational efficiency outcomes.
For HD and $F_1$ score, the CT algorithm generally outperformed all other methods regardless of $\alpha$ and $d$, with an average $F_1$ score 0.901 across all conditions.
The accuracy of the CT algorithm slightly decreases as $\alpha$ increases with a notable drop in accuracy when $\alpha = 1$.
This pattern matches the trends in thresholdability and sortability of the correlations which are displayed in Table~\ref{tab:sortability}.
Notably, even if a lack of thresholdability precludes perfect model recovery, a near-perfect model can be recovered if the parameters are near-thresholdable, as indicated by a high sortability score.
For the EFA methods, for $d$ ranging from 3 to 5, these methods were generally not affected by $\alpha$, and had overall middling accuracies, which ranged from 0.431 to 0.687.
At $d = 2$, there was an exception to these patterns, where the CT algorithm had a slight increase in accuracy at $\alpha = 1$.
This is due to the fact that $d$ was overestimated for $\alpha < 1$.
Close inspection of these cases showed that a third factor was extracted under these conditions about half the time, which was highly correlated with the original two (as identified by their unique children).
This third factor was able to absorb multiple paths from the original two, leading to comparable fit and ultimately less parameters, which may explain its selection by BIC.
We provide a concrete example in Appendix~\ref{app:low_dim}.
Returning to the overall results, the EFA-LASSO and EFA-MCP methods simply decreased in accuracy as $\alpha$ increased, when $d = 2$.
For the number of latent factors, the CT algorithm was generally accurate except when $d = 2$ as observed earlier, where it over-estimated with a $\hat{d}$ with an average of about 2.5, and decreased until it achieved an accurate $\hat{d}$ of 2 at $\alpha = 1$.
The EFA method generally estimated the correct number of factors across all $d$ and $\alpha$, and the EFA-LASSO and EFA-MCP methods generally estimated the number of latent factors correctly at low values of $\alpha$, but underestimated as $\alpha$ increased.

\begin{table}[t]
\tbl{Proportion of datasets that were threshsoldable per $d$ and $\alpha$, and average sortability proportions per $d$ and $\alpha$.}
{\begin{tabular}{cccccccccccc}
\toprule
& & \multicolumn{5}{c}{Thresholdable} & \multicolumn{5}{c}{Sortability}\\
\cmidrule(lr){3-7}\cmidrule(lr){8-12}
\multirow{2}{*}{Correlation} & \multirow{2}{*}{$d$} & \multicolumn{5}{c}{$\alpha$} & \multicolumn{5}{c}{$\alpha$}\\
\cmidrule(lr){3-7}\cmidrule(lr){8-12}
& & 0 & 0.25 & 0.50 & 0.75 & 1.00 & 0 & 0.25 & 0.50 & 0.75 & 1.00\\
\midrule
\multirow{4}{*}{Population} & 2 & 1.000 & 1.000 & 1.000 & 1.000 & 0.950 & 1.000 & 1.000 & 1.000 & 1.000 & 0.995\\
& 3 & 1.000 & 1.000 & 1.000 & 0.970 & 0.070 & 1.000 & 1.000 & 1.000 & 0.999 & 0.729\\
& 4 & 1.000 & 1.000 & 1.000 & 0.510 & 0.000 & 1.000 & 1.000 & 1.000 & 0.978 & 0.650\\
& 5 & 1.000 & 1.000 & 1.000 & 0.190 & 0.000 & 1.000 & 1.000 & 1.000 & 0.958 & 0.587\\
\midrule
\multirow{4}{*}{Sample} & 2 & 1.000 & 1.000 & 1.000 & 1.000 & 0.820 & 1.000 & 1.000 & 1.000 & 1.000 & 0.978\\
& 3 & 1.000 & 1.000 & 1.000 & 0.760 & 0.020 & 1.000 & 1.000 & 1.000 & 0.986 & 0.660\\
& 4 & 1.000 & 1.000 & 0.950 & 0.120 & 0.000 & 1.000 & 1.000 & 0.999 & 0.927 & 0.527\\
& 5 & 1.000 & 1.000 & 0.690 & 0.030 & 0.000 & 1.000 & 1.000 & 0.995 & 0.876 & 0.475\\
\bottomrule
\end{tabular}}
\label{tab:sortability}
\end{table}

For the fit measures, the test log-likelihood, TLI, and RMSEA all agree in the patterns of results.
Generally speaking, the CT algorithm out-performed all other methods, with a few exceptions at $\alpha = 1$, where it was then out-performed by EFA.
Across all conditions, the fit gets worse as $\alpha$ increases.
EFA-LASSO and EFA-MCP were affected by this the most, and were more sensitive to this at lower $d$.
Standard EFA was the most robust of the EFA methods, performing similarly to the CT algorithm except at the aformentioned $\alpha = 1$ condition.
On the other hand, the CT algorithm and EFA methods start to see the pattern of declining model fitting only at $\alpha = 1$.
While there does seem to be a slightly curved relationship between fit and $\alpha$, this is attributable to several outliers downardly influencing the average at lower $\alpha$.
These can be seen in our boxplots in Appendix~\ref{app:additional_plots}.
Moreover, these boxplots showed that EFA-LASSO and EFA-MCP had larger variability in their fit measures in the $\alpha \geq 0.5$ conditions.
For the number of models tested per method, the results are presented on the $\log_{10}$ scale.
The number of models is not very much affected by $d$ or $\alpha$, and the rank order from the least to most number of models is EFA, CT, EFA-LASSO, and EFA-MCP.

Overall, these results show that the CT algorithm maintains high structural accuracy and statistical fit across a broad range of conditions, with performance declining only under extreme violations of thresholdability ($\alpha = 1$).
Importantly, it achieved this while evaluating substantially far fewer models than the penalized EFA methods.
Performance declines closely followed reductions in sortability, suggesting that even when thresholdability fails, the algorithm can still perform well provided sortability remains moderately high.

\subsection{Violations to the Unique Child Condition} \label{sec:ucc_sim}

For violations to the unique child condition, we used the same low-dimensional setting as the thresholdability simulation in Section~\ref{sec:thresh_sim}: $d \in \{2, 3, 4, 5\}$  with the number of children per latent variable set to 5, $n \in \{250, 500, 1000\}$, and 100 datasets per condition.
Once again, we will only examine the results for the $n = 1000$ case, since the $n = 250$ and $n = 500$ conditions similar patterns of results, but these are provided in the supplementary materials, linked in Appendix~\ref{app:osf}.

To violate the unique child condition, we began with an independent cluster structure for $\Lambda$, but then randomly selected a proportion, $\beta \in [0, 1]$, of latent variables for which the unique child condition would be violated.
For each of these latent variables, all their children would be given cross-loadings to another latent variable randomly.
All possible values of $\beta$ were tested, according to how many latent variables were used in the condition (e.g., if $d = 2$, then $\beta \in \{0, 0.5, 1\}$).
Thresholdability was enforced by setting $\Phi = I$.

\begin{figure}[tb]
  \centering
  \includegraphics[width=\textwidth, keepaspectratio]{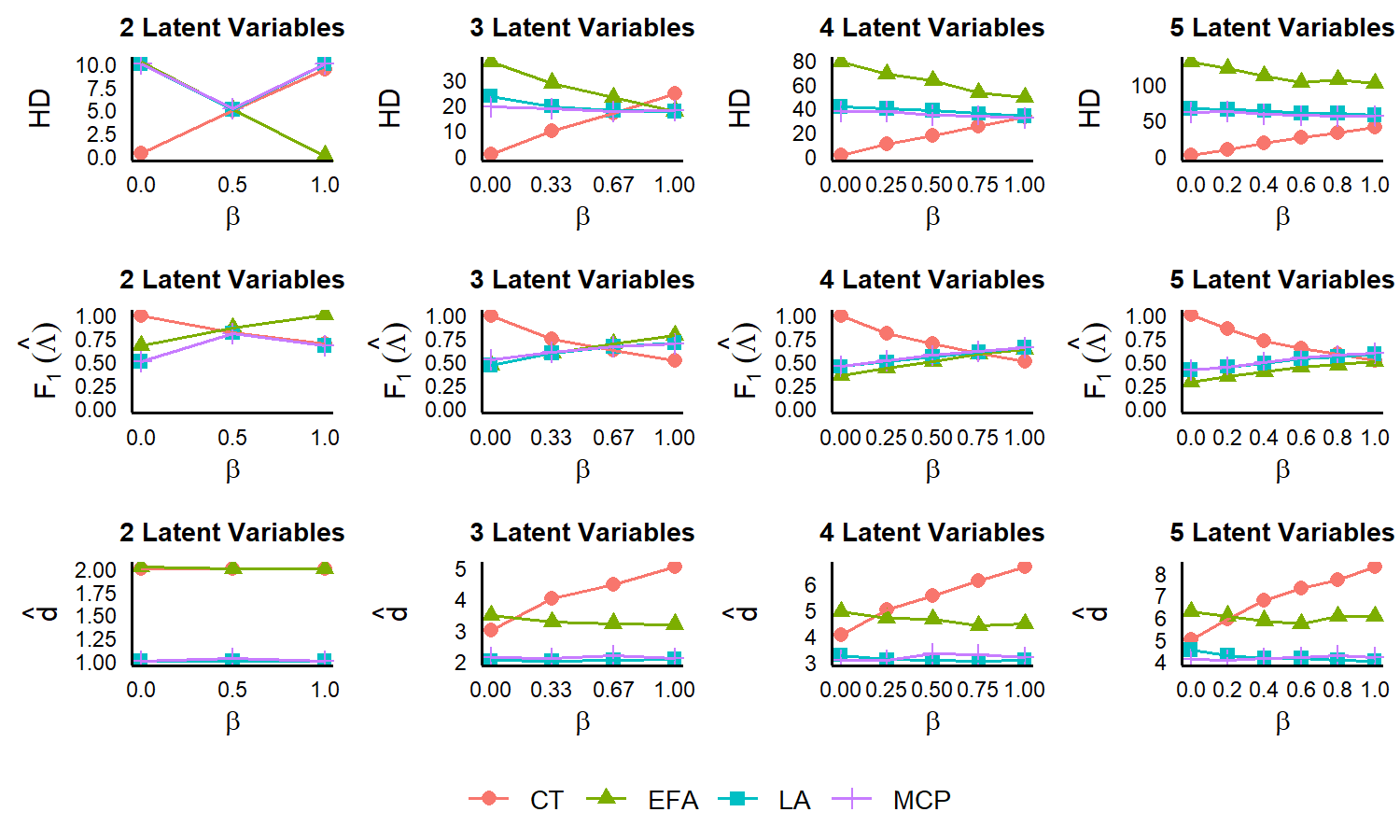}
  \caption{Average structural accuracy statistics for the low-dimensional simulation, as a function of $\beta$.
  For $d = 2$, we note that CT and EFA are largely overlapping on $\hat{d}$, and LA and MCP are largely overlapping on all three metrics.}
  \label{fig:beta_str_1000}
\end{figure}

\begin{figure}[tb]
  \centering
  \includegraphics[width=\textwidth, keepaspectratio]{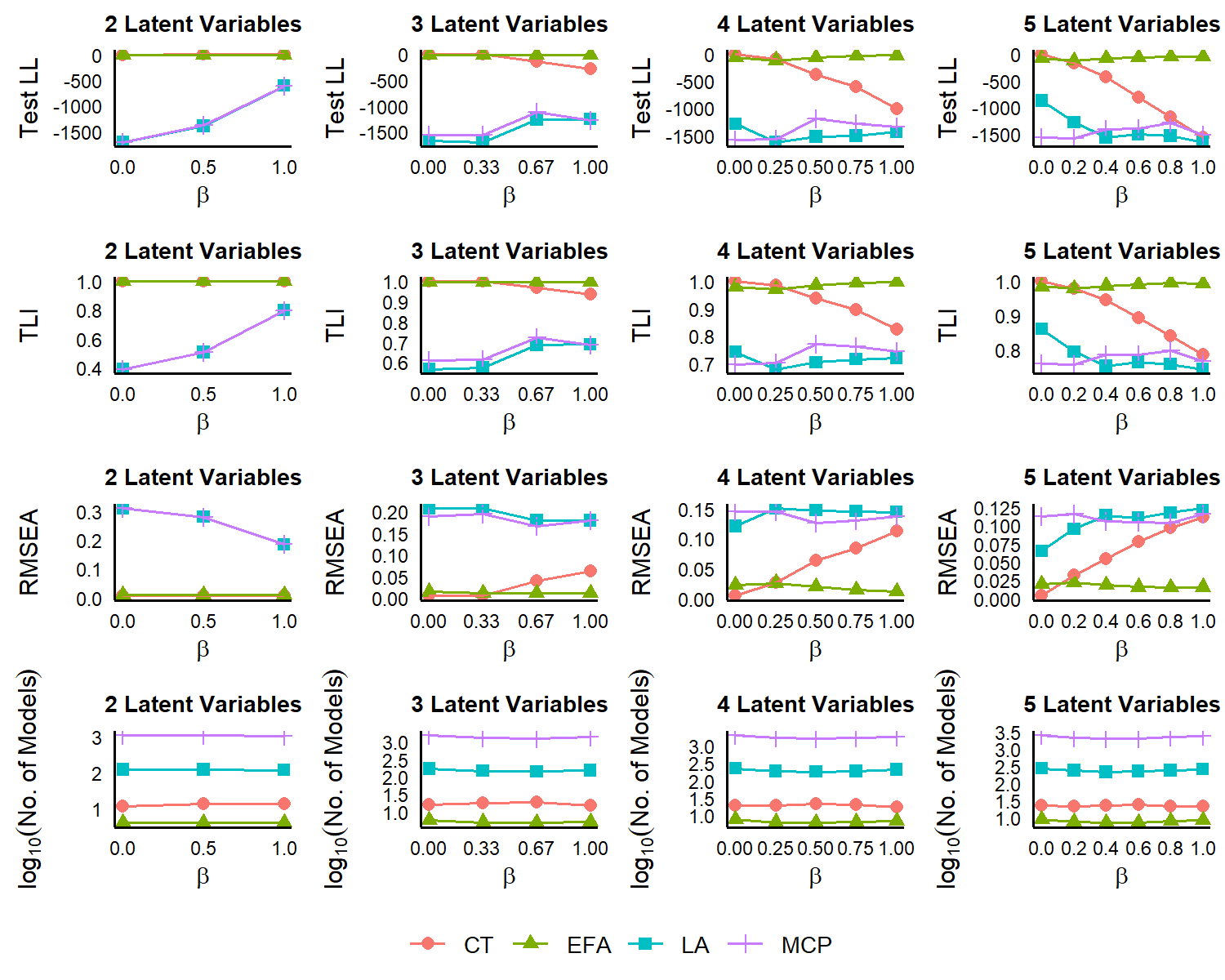}
  \caption{Average model fit and computational efficiency statistics for the low-dimensional simulation, as a function of $\beta$.}
  \label{fig:beta_fit_1000}
\end{figure}

The results for varying $\beta$ are displayed in Figure~\ref{fig:beta_str_1000} for the structural accuracy outcomes and in Figure~\ref{fig:beta_fit_1000} for the statistical fit and computational efficiency outcomes.
For the $F_1$ score, all three EFA methods display low to moderate accuracy, increasing slightly as $\beta$ increased.
In contrast, $F_1$ score for the CT algorithm decreased with $\beta$, but ultimately does not perform much worse than the EFA methods even when $\beta = 1$.
This decrease may be attributable to the fact that the CT algorithm tended to over-extract factors as $\beta$ increased.
In contrast, the EFA-LASSO and EFA-MCP methods tended to under-extract factors and EFA over-extracted factors regardless of the level of $\beta$.

\begin{figure}[t!]
  \centering
  \includegraphics[width=0.4\textwidth, keepaspectratio]{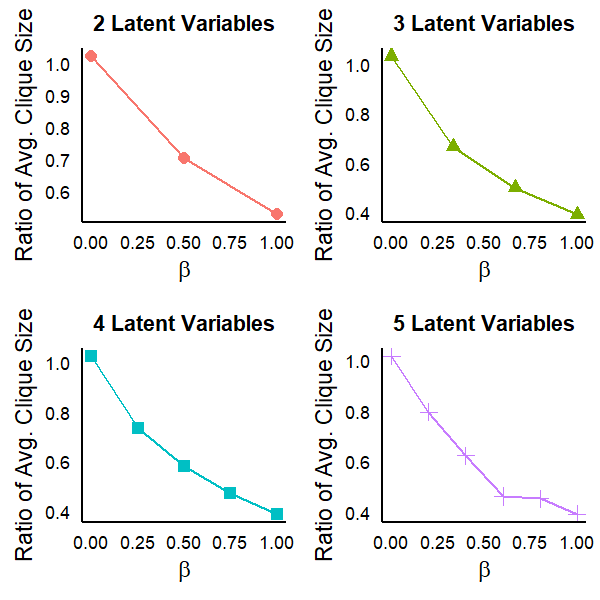}
  \caption{Ratios between the average maximal clique size of the estimated model's graph ($\lvert \widehat{C_k} \rvert$) and the true model's graph ($\lvert C_k \rvert$), calculated per $d$ and $\beta$.}
  \label{fig:clique_size}
\end{figure}

The overestimation of $d$ by the CT algorithm may be explained by a factor splitting phenomenon due to the CT algorithm's use of maximal cliques to identify latent variables.
That is, the effect of $\beta$ is essentially that more latent variables are no longer identified by their unique child sets.
This leads to the non-identified latent variables splitting their children over multiple smaller maximal cliques in order to explain the correlation pattern.
In Figure~\ref{fig:clique_size}, we show the ratios between the average maximal clique size of the estimated model's graph ($\lvert \widehat{C_k} \rvert$) and the true model's graph ($\lvert C_k \rvert$).
We can see that the ratio starts at about 1 when $\beta = 0$, indicating correctly estimated clique sizes when the UCC holds.
However, when $\beta$ increases, this ratio decreases immediately across all $d$.
This indicates that the average clique size in the estimated graph becomes smaller than that of the true graph.
We note that the candidate set of structures included solutions with $\hat{d}$ and $F_1$ scores closer to the true model than the one selected by BIC.
This indicates that BIC favored models with increased $d$ due to improved fit outweighing the complexity penalty, even when more accurate structures were available.

As with $\alpha$, all three fit measures agree on their pattern of results.
EFA displayed the best fit, which may be attributable to the fact that many extra factor loadings are added to violate the unique child condition.
All these extra paths make the true model closer to a saturated factor loading matrix which is assumed by EFA.
For the CT algorithm, the fit is very good when $\beta = 0$, but decreases as $\beta$ increases, and this effect gets stronger as $d$ increases.
Overall, the fit of the CT algorithm is good to acceptable for $\beta < 0.5$,
however beyond this, the CT algorithm's robustness diminishes.
The EFA-LASSO and EFA-MCP methods have poorer fit regardless of $\beta$, and generally does not do better than the CT algorithm even in the worst case of $\beta = 1$.
For the number of models, the rank order is identical to that of the $\alpha$ condition, with the least to most number of models being EFA, CT, EFA-LASSO, and EFA-MCP, and this does not change with $\beta$.

\subsection{High-Dimensional Settings} \label{sec:highdim}

Our high-dimensional simulation dealt with the situation of $n < p$.
Specifically, we increased both $p$ and $d$ as a function of $n$ by setting $p = 1.5n$ and $d = 0.1n$, for $n \in \{250, 500, 1000\}$.
This set of simulation settings will also showcase the performance of the CT algorithm on data with a large number of latent factors $d\in\{25,50,100\}$.
Computing the MLE when $n < p$ with large $p$ is prohibitively slow, so we focused on only studying the structural accuracy of the CT algorithm.
We did this by running the CT algorithm except for the estimation and model selection steps.
Doing this yields a set of proposed structures, from which we examined the structure closest to the true structure.
As with the low-dimensional settings, we varied $\alpha$ and $\beta$, varying them across $\{0, 0.25, 0.5, 0.75, 1\}$ for each.
For the cutoffs $\tau_k$, 50 equidistant points from 0 to 1 were used.
We collected HD, the $F_1$ score of $\hat{\Lambda}$, and $\hat{d}$ for structural accuracy, and examined the computation time for computational efficiency.

\begin{figure}[tb]
  \centering
  \includegraphics[width=\textwidth, keepaspectratio]{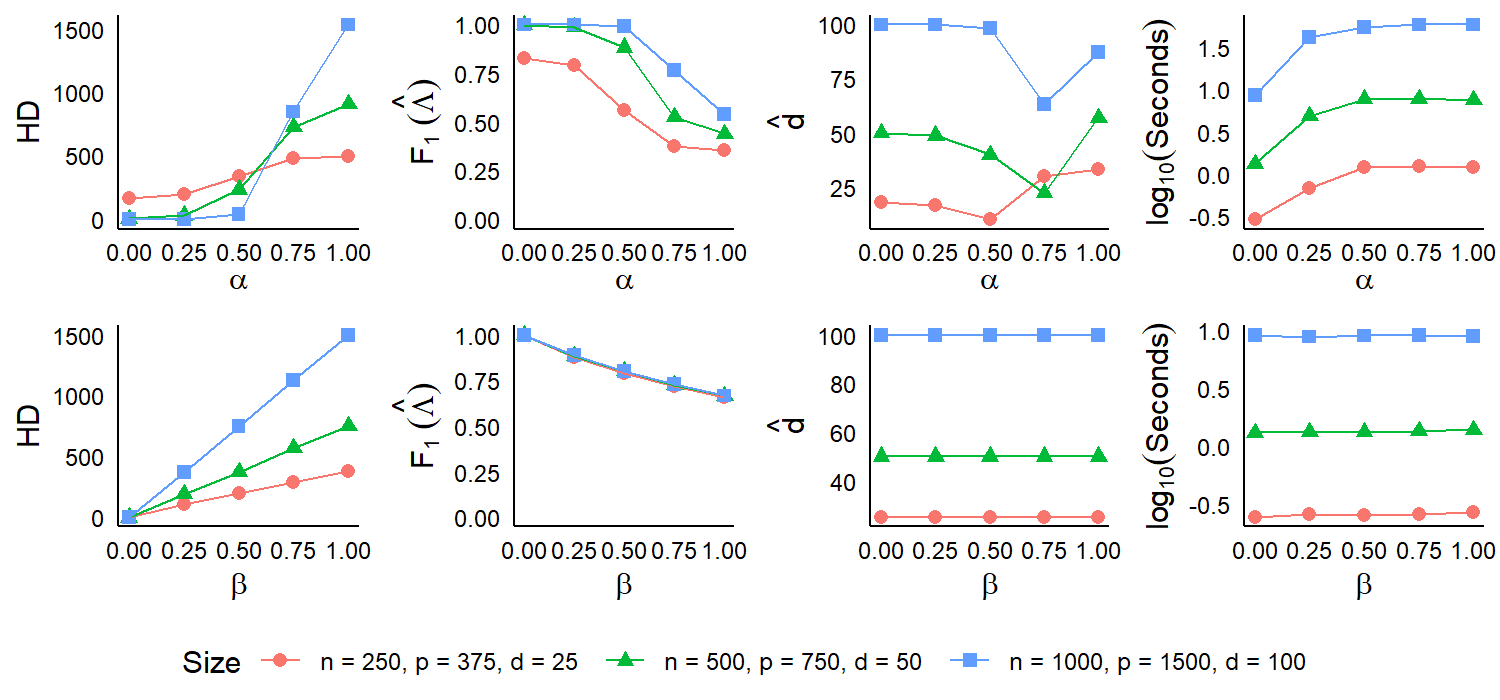}
  \caption{Average structural accuracy and computational efficiency statistics for the high-dimensional simulation, as a function of $\alpha$ and $\beta$.}
  \label{fig:hd}
\end{figure}

The results of this simulation are displayed in Figure~\ref{fig:hd}.
With respect to increases in $\alpha$, HD and $F_1$ score are adversely affected across all sizes.
However, this effect diminishes at higher sample sizes, indicating that sample size increases robustness despite a proportional increase in $p$ and $d$.
For $\hat{d}$, the trend is non-monotonic.
The average starts out fairly accurate with a slightly decreasing trend, then increases again after a certain point.
This pattern is observed in all three conditions, albeit at varying levels of $\alpha$.
This may be explained by noting that $\hat{d}$ and $\tau$ do not have a monotone relationship.
For example, as $\tau$ moves from 1 to 0, maximal cliques can get added (thereby increasing $\hat{d}$), but it can also cause maximal cliques to merge (thereby decreasing $\hat{d}$).
Additionally, when $\alpha$ increases, the entire landscape of correlations will change with respect to $\tau$, thus groups of variables that may have formed cliques may no longer do so.
This has the effect of slowly changing the best $F_1(\hat{\Lambda})$ solution, which explains the slow decrease in $\hat{d}$.
Then once $\alpha$ has increased to the point which that solution (or a similar one) is no longer the closest, there may be a sudden jump in $\hat{d}$ moving into a new solution.
We provide an illustrated example of this effect in Figure~\ref{fig:ad_hoc}.
Returning to the remainder of the results, the computation time increases with both $\alpha$ and model size, though there is no apparent interaction.

With respect to increases in $\beta$, HD and $F_1$ score are affected linearly.
Of note, the $F_1$ score is nearly identical across all three model sizes, suggesting a robust performance with respect to increase in the dimension of the model and the data size.
Unlike the $\alpha$ settings however, $\hat{d}$ is stable and accurate across both $\beta$ and model sizes, and the same pattern holds for computation time. 

\section{Real Data Example}

For a real data example, we analyzed measures of youth impulsivity from the Adolescent Brain Cognitive Development (ABCD) Study (\url{https://abcdstudy.org}).
The ABCD is a large, multi-site study, whose data are publicly available \citep{Volkow2018}.
The data collection was approved by the institutional review boards of all the participating sites \citep{Clark2018}.
To reduce the sample size to a more realistic scale, we selected one site at random for analysis ($n = 582$).

\begin{figure}[tb]
  \centering
  \includegraphics[width=\textwidth, keepaspectratio]{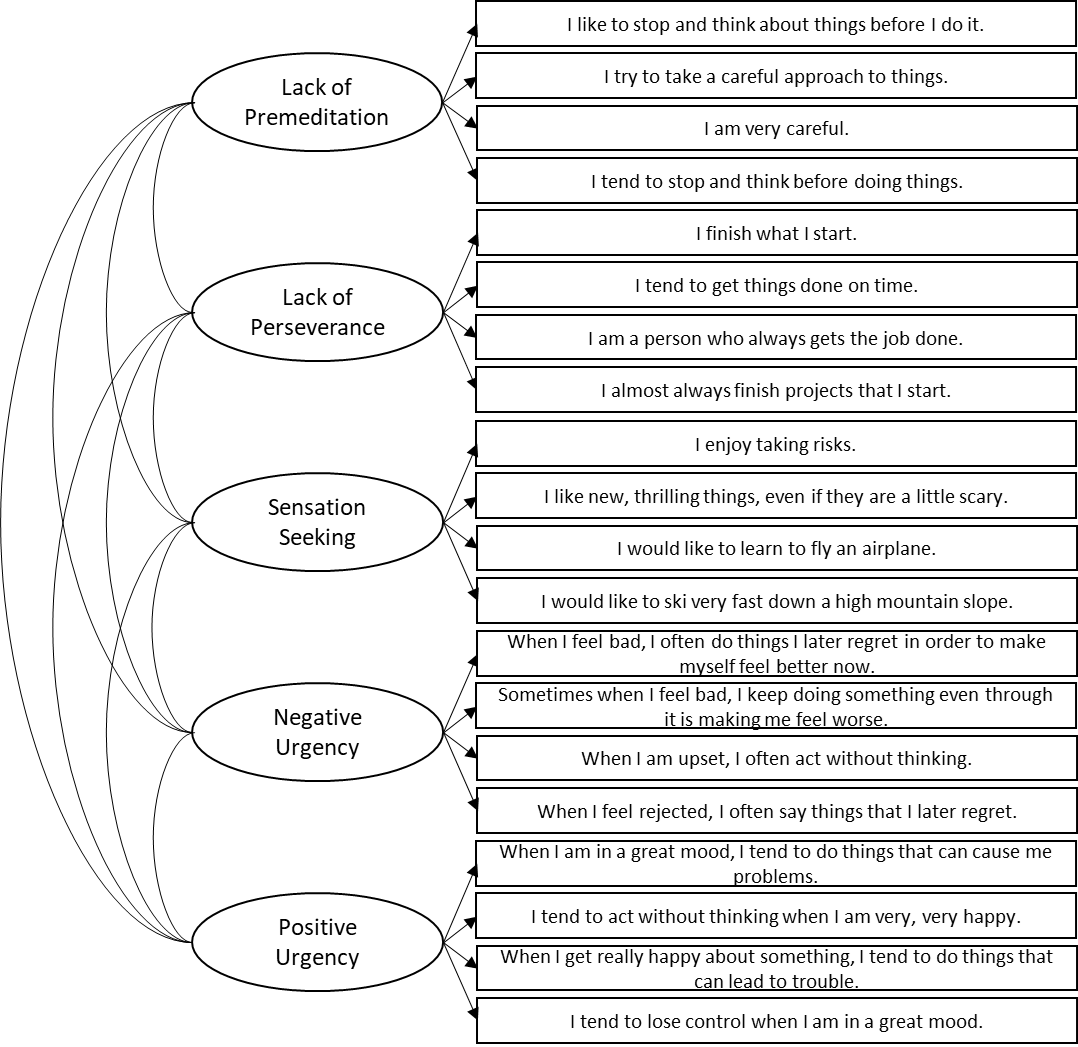}
  \caption{Path diagram of the youth UPPS-P scale.}
  \label{fig:upps}
\end{figure}

We analyzed the youth version of the UPPS-P scale.
The UPPS-P is a five factor personality scale designed to measure impulsivity, and has been argued to play a prominent role in various forms of psychopathology \citep{Whiteside2001}.
It is hypothesized to follow an independent cluster structure, with the youth-specific version having four observed variables for each of the five latent variables \citep{Watts2020}.
These latent factors represent (a lack of) premeditation, (a lack of) perseverance, sensation seeking, negative urgency, and positive urgency.
A path diagram of this model is illustrated in Figure~\ref{fig:upps}, along with the scale items.
We followed the analysis procedure and outcomes of the low-dimensional simulations described in Section~\ref{sec:metrics}.
The hypothesized model of Figure~\ref{fig:upps} was taken to be the true model for the purposes of calculating HDs and $F_1$ scores.

The results are displayed in Table~\ref{tab:rd_results}.
The CT algorithm performed the best in terms of obtaining a structure closest to the hypothesized model by a large margin.
It obtained an $F_1$ score of 0.974, which resulted from an HD of 1, only omitting a single factor loading.
This factor loading was by far the weakest loading ($\lambda = 0.178$), which was between the item ``I would like to learn to fly an airplane'' and the sensation seeking construct.
The next closest was EFA-MCP, with an $F_1$ score of 0.561, followed by EFA (0.320) and EFA-LASSO (0.281).
Notably, the CT algorithm was the only method that obtained the hypothesized number of factors, whereas the rest of the methods under-estimated by extracting 2 or 4 factors.
All the fit measures followed the same patterns as the $F_1$ score, although the differences were not as dramatic.
In terms of computational efficiency, the CT algorithm obtained the closest structure by only checking 24 models, in contrast to the next closest model found by EFA-MCP, which was found by checking 1350 models.

\begin{table}[t]
\tbl{Structural accuracy, model fit, and number of models per method for the real data analysis.}
{\begin{tabular}{ccccSccS}
\toprule
Method & HD & $F_1(\hat{\Lambda})$ & $\hat{d}$ & {BIC} & TLI & RMSEA & {Models}\\
\midrule
Hypothesized Model &  0 & 1.000 & 5 &  9686.800 & 0.889 & 0.050 & 1\\
CT                 &  1 & 0.974 & 5 &  9689.250 & 0.886 & 0.050 & 24\\
EFA-LASSO          & 41 & 0.281 & 2 & 10167.550 & 0.660 & 0.087 & 150\\
EFA-MCP            & 25 & 0.561 & 4 &  9896.140 & 0.806 & 0.066 & 1350\\
EFA                & 68 & 0.320 & 4 & 10031.750 & 0.796 & 0.067 & 5\\
\bottomrule
\end{tabular}}
\label{tab:rd_results}
\end{table}

\section{Conclusions}

In this research, we introduced the CT algorithm and examined its ability to discern factor analysis structures.
Previous research has shown that the thresholdability and unique child assumptions are required for consistent estimates in terms of both the structures and parameter values.
As such, we tested the robustness to these assumptions in both low-dimensional and high-dimensional settings, and compared it to other structure learning methods.
We also provided a comparison of methods in a real data example.

We showed that the CT algorithm performed well when its assumptions were met, and showed good robustness to violations of thresholdability and moderate robustness to violations of the unique child condition.
Interestingly, the penalized EFA methods also showed sensitivity to violating these assumptions, in which the CT algorithm generally outperformed these methods or were on par with them.
This shows that the CT algorithm can be used as a competitive method for EFA, despite being developed under these assumptions.
The CT algorithm also showed viability as a structure learning method in high-dimensional settings, where penalized EFA methods are impractical due to their computational intensity.
Moreover, regardless of the condition, the CT algorithm has vastly superior computational efficiency, as it checks much fewer candidate models.
These performance patterns were also corroborated in the real data example.

For future research, the CT algorithm can be generalized in a few ways.
Since the CT algorithm only requires a correlation matrix for structure learning, it would be straightforward to generalize this to non-linear factor analysis models by using non-linear correlation metrics (e.g., a polychoric correlation matrix for IRT structures).
Moreover, even if $\theta$ is thresholdable in the population, this may be hampered by variability in the sample.
This may be accounted for by incorporating subsampling or bootstrap aggregation techniques (e.g., \citealt{Meinshausen_2010}).

\section*{Funding}

This work was supported by NSF grant DMS-2305631.

\newpage
\bibliographystyle{apacite}
\bibliography{references}

\newpage
\appendix

\section{Supplementary Materials} \label{app:osf}

Supplementary plots, code, and data for this study are available at:\\ \url{https://osf.io/4m2pv/?view\_only=7e0bb09f31924fc4b05d053af978165d}

\section{Additional Simulation Details}

\subsection{Data Generation Details} \label{app:data_gen}

The non-zero entries of $\Lambda$ were generated with a uniform distribution on the range $[0.6, 0.8]$
To generate $\Phi$, we began by creating a random $d \times d$ matrix $A$ whose entries were drawn from the standard uniform distribution.
We then took the product $A^T A$ and scaled the off-diagonals to the range $[0.6, 0.8]$, and set the diagonals to 1, which guaranteed positive definiteness.
Then we took $\Phi$ as this re-scaled product matrix.

\subsection{Outcomes} \label{app:outcomes}

\subsubsection{Structural Accuracy} \label{app:struc_acc}

For the simulation studies we studied Hamming distance and $F_1$ score of $\hat{\Lambda}$.
To reconcile the fact that $\hat{\Lambda}$ and $\Lambda$ may not have the same number of columns, we augmented the smaller of the two matrices with columns of zeroes until their columns matched.
Let us define these possibly augmented versions as $\hat{\Lambda}^*$ and $\Lambda^*$.
Additionally, it may be the case that the column-order of $\hat{\Lambda}^*$ and $\Lambda^*$ may not match.
Hence, we computed the minimum Hamming distance among all column permutation of $\hat{\Lambda}^*$.
That is, we define a HD as
\begin{equation} \label{eq:hd}
\text{HD} \coloneqq \min_{P} \left[ \lvert \mathcal{A}(\hat{\Lambda}^* P) \, \triangle \, \mathcal{A}(\Lambda^*) \rvert \right],
\end{equation}
where $\triangle$ is the symmetric difference or disjunctive union between two sets, and $P$ is a permutation matrix.
The $F_1$ score was calculated based on this permutation.

\subsubsection{Sortability} \label{app:sortability}

Let the set of correlation magnitudes among observed variables that share parents be $R_{\text{SP}} \coloneqq \{\lvert r_{ij} \rvert : (i, j) \in E_0\}$ and the set of correlation magnitudes among observed variables that do not share parents be $R^c_{\text{SP}} \coloneqq \{\lvert r_{kl} \rvert : (k, l) \in E^c_0\}$, where $r_{ij}$ are the sample correlations among $X_i$ and $X_j$.
Then we define the sample \textit{sortability} as the proportion
\begin{equation}
\begin{aligned}
\dfrac{\lvert \{(i,j) : (i, j) \in E_0, \lvert r_{ij} \rvert > \max R^c_{\text{SP}} \} \cup \{(k,l) : (k, l) \in E^c_0, \lvert r_{kl} \rvert < \min R_{\text{SP}} \}\rvert}{p(p -1)/2}.
\end{aligned}
\end{equation}
The population sortability is analagously defined simply by using $\rho_{ij}$ in place of $r_{ij}$.

\newpage

\section{Additional Analyses}

\subsection{Example Solution in Low-Dimensional Setting} \label{app:low_dim}

We inspect an example when $d = 2$ and $\alpha = 0$ in the low-dimension setting.
Consider the following $\hat{\Lambda}$ matrices given by the MLE (which uses the true structure) and the CT algorithm:

\begin{equation}
\hat{\Lambda}_{\text{MLE}} = \bbm
0.849 & \\
& 0.860 \\
0.611 & 0.611 \\
0.860 & \\
0.583 & 0.607 \\
0.811 & \\
& 0.837 \\
0.586 & 0.614 \\
0.551 & 0.562 \\
0.543 & 0.574 \\
\ebm,\,\,
\hat{\Lambda}_{\text{CT}} = \bbm
0.848  & & \\
& \phantom{-}0.857 & \\
& -0.026 & 0.882 \\
0.860 & & \\
& & 0.841 \\
0.810 & & \\
& \phantom{-}0.836 & \\
& & 0.847 \\
& & 0.786 \\
& & 0.789
\ebm,
\end{equation}
where the blank entries are zero.
Also consider their corresponding $\Phi$ matrices:
\begin{equation}
\hat{\Phi}_{\text{MLE}} = \bbm
\phantom{-}1.000 & -0.001\\
-0.001  & \phantom{-}1.000
\ebm,\,\,
\hat{\Phi}_{\text{CT}} = \bbm
\phantom{-}1.000 & -0.003 & 0.694\\
-0.003 & \phantom{-}1.000 & 0.725\\
\phantom{-}0.694 & \phantom{-}0.725 & 1.000
\ebm.
\end{equation}
The BIC for these solutions are 2841.686 and 2827.482 for the MLE and CT algorithm, respectively.
This shows that for very small $d$ is it possible to more parsimoniously explain the correlation pattern by adding a latent factor.
In this example, it appears that the third factor absorbed the paths from $L_1$ and $L_2$ into $X_5$, $X_9$, $X_8$, and $X_{10}$, ultimately using two less parameters than the MLE solution.

\newpage

\subsection{Example Solution in High-Dimensional Setting} \label{app:high_dim}

\begin{figure}[!h]
  \centering
  \includegraphics[width=\textwidth, keepaspectratio]{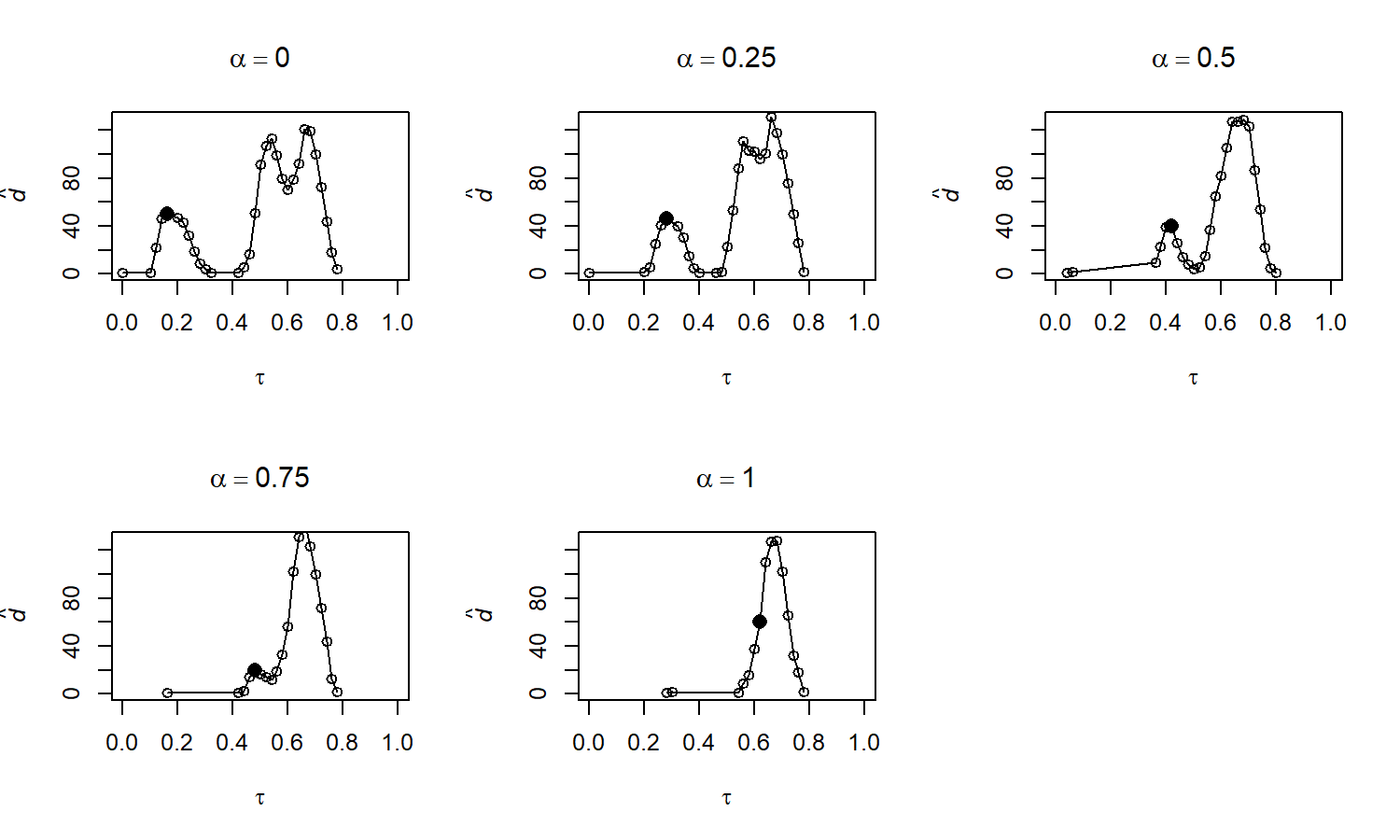}
  \caption{Examples of a high dimensional structure proposed by the CT algorithm in the $n = 500$, $p = 750$, $d = 50$ case, where we can see the $\hat{d}$ per $\tau$. Each open dot represents a proposed structure, and the solid dot represents the structure with the best $F_1(\hat{\Lambda})$ score. The landscape of possible $\hat{d}$ changes as $\alpha$ increases, where a critical change occurs between $\alpha = 0.75$ and $\alpha = 1$ causing a jump in $\hat{d}$ for the best $F_1(\hat{\Lambda})$ solution.}
  \label{fig:ad_hoc}
\end{figure}

\newpage

\section{Additional Plots} \label{app:additional_plots}

\begin{figure}[h!]
  \centering
  \includegraphics[width=.9\textwidth, keepaspectratio]{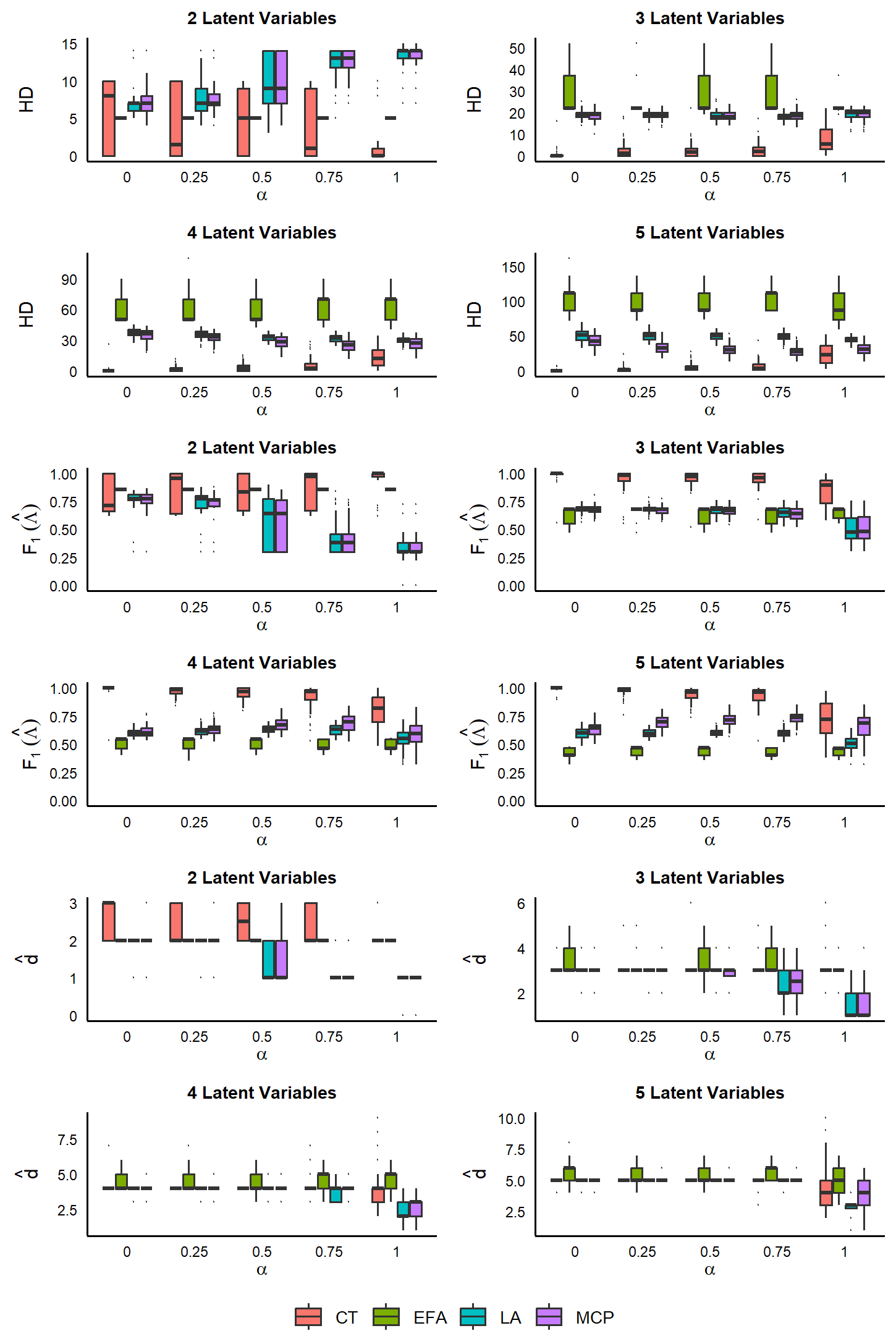}
  \caption{Boxplots of structural accuracy statistics for the low-dimensional simulation when $n = 1000$, as a function of $\alpha$.}
\end{figure}

\begin{figure}[h!]
  \centering
  \includegraphics[width=.9\textwidth, keepaspectratio]{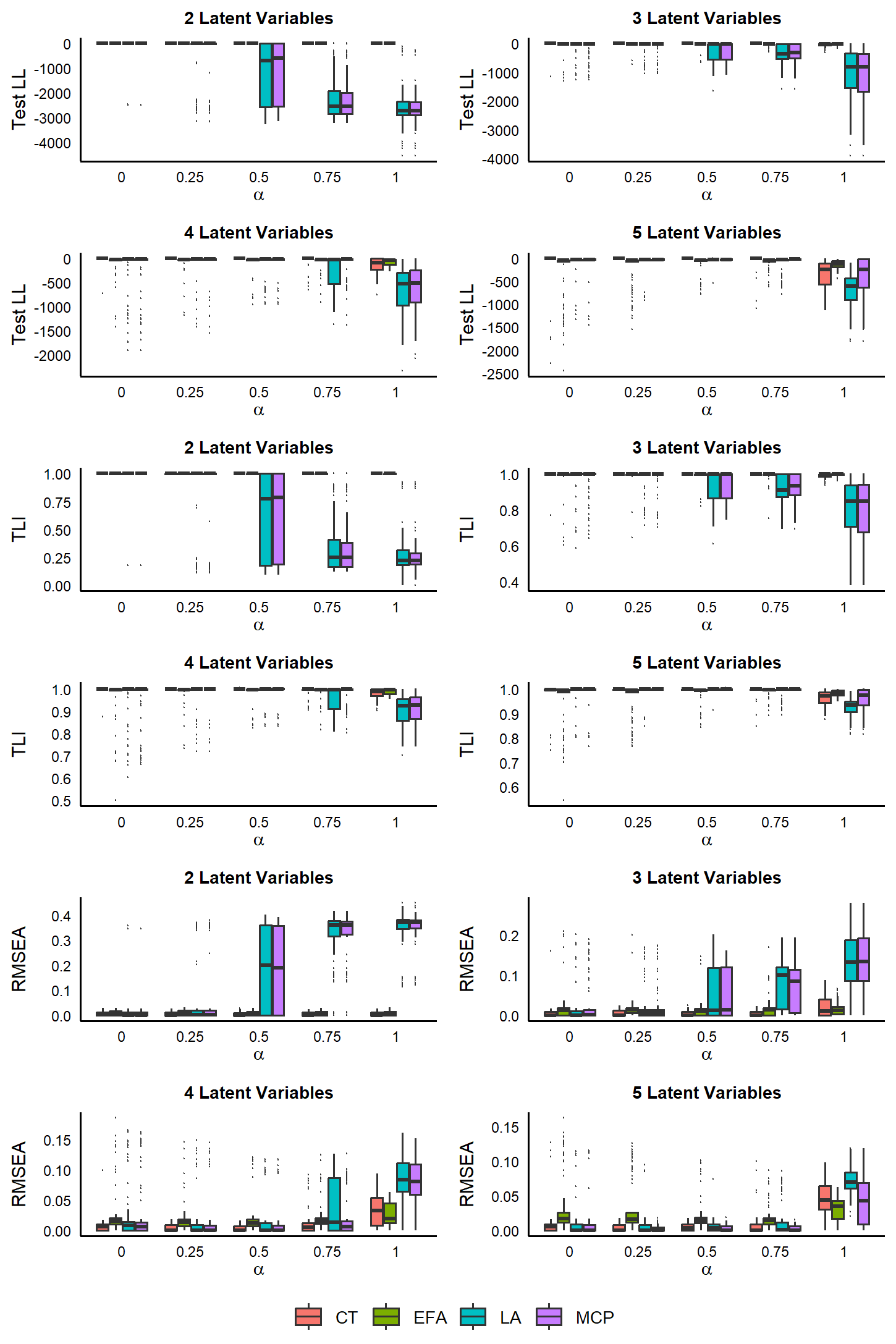}
  \caption{Boxplots of fit statistics for the low-dimensional simulation when $n = 1000$, as a function of $\alpha$.}
\end{figure}

\begin{figure}[h!]
  \centering
  \includegraphics[width=.9\textwidth, keepaspectratio]{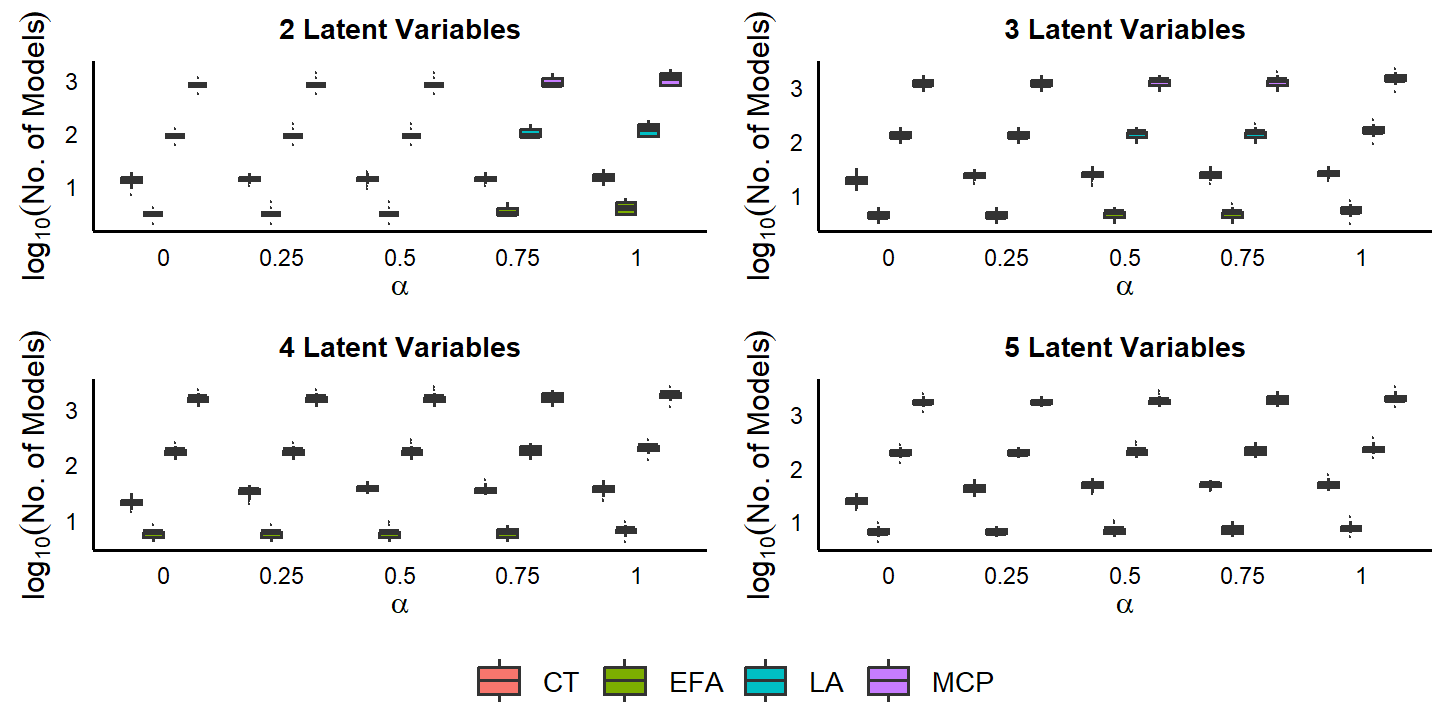}
  \caption{Boxplots of computational efficiency statistics for the low-dimensional simulation when $n = 1000$, as a function of $\alpha$.}
\end{figure}

\begin{figure}[h!]
  \centering
  \includegraphics[width=.9\textwidth, keepaspectratio]{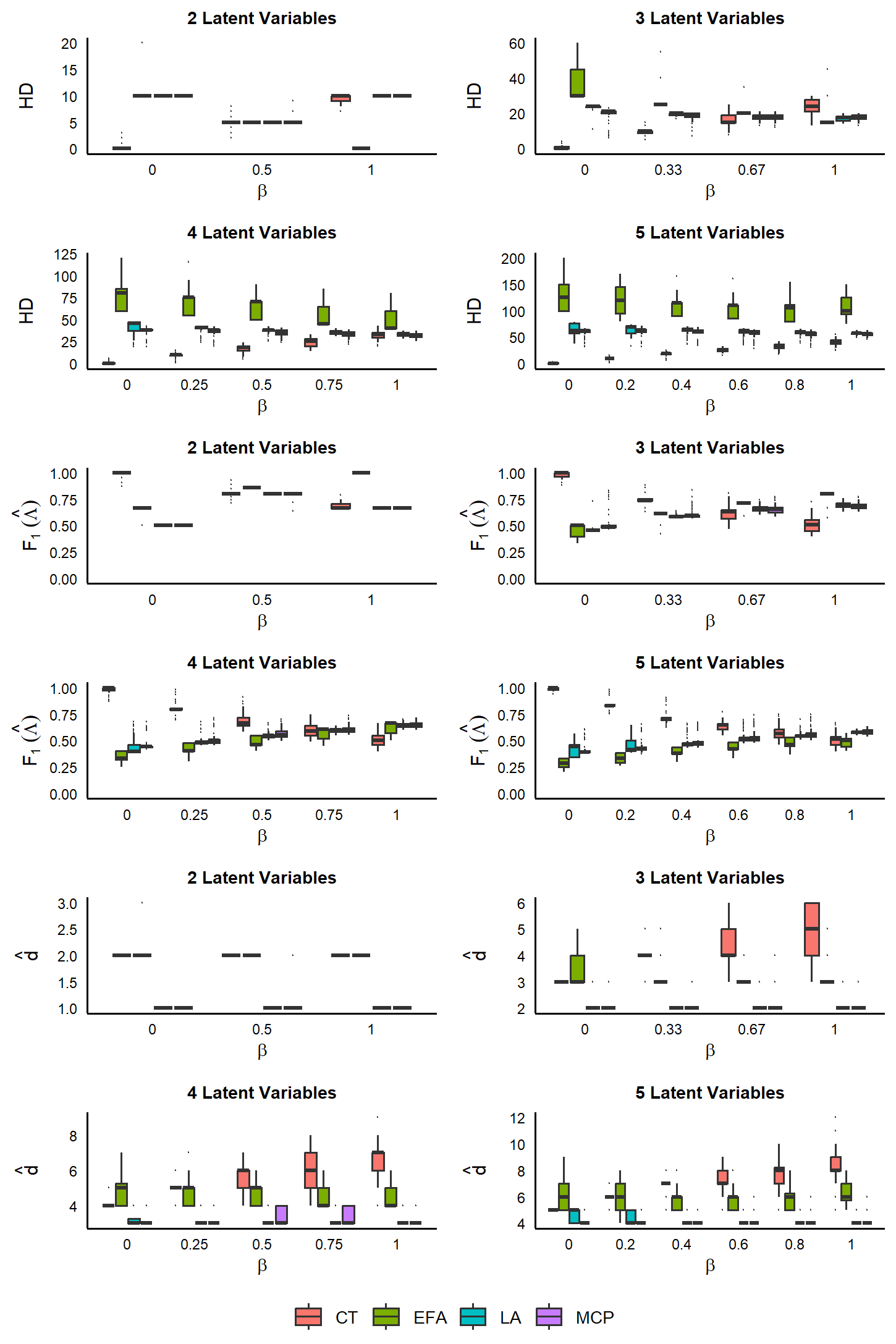}
  \caption{Boxplots of structural accuracy statistics for the low-dimensional simulation when $n = 1000$, as a function of $\beta$.}
\end{figure}

\begin{figure}[h!]
  \centering
  \includegraphics[width=.9\textwidth, keepaspectratio]{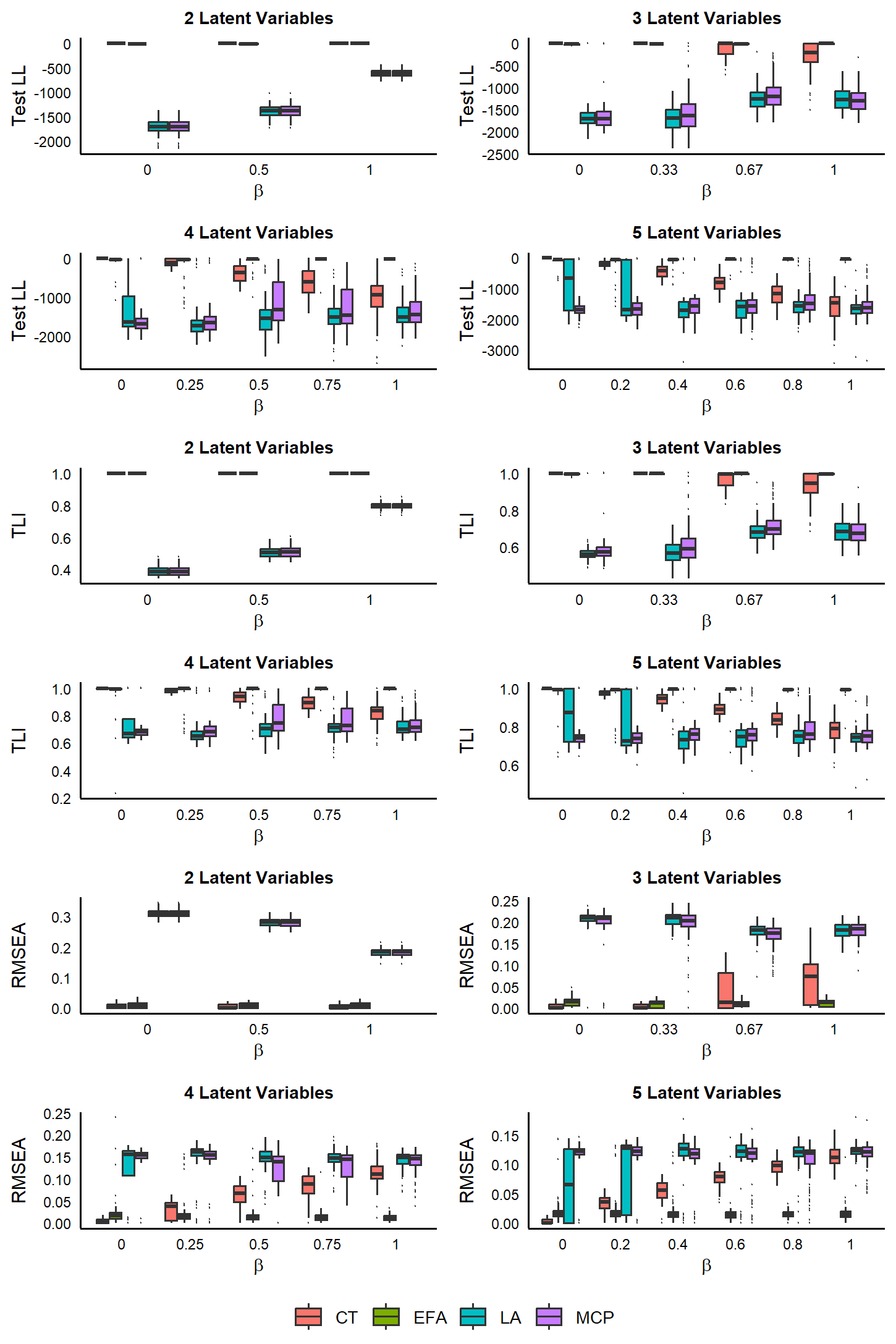}
  \caption{Boxplots of fit statistics for the low-dimensional simulation when $n = 1000$, as a function of $\beta$.}
\end{figure}

\begin{figure}[h!]
  \centering
  \includegraphics[width=.9\textwidth, keepaspectratio]{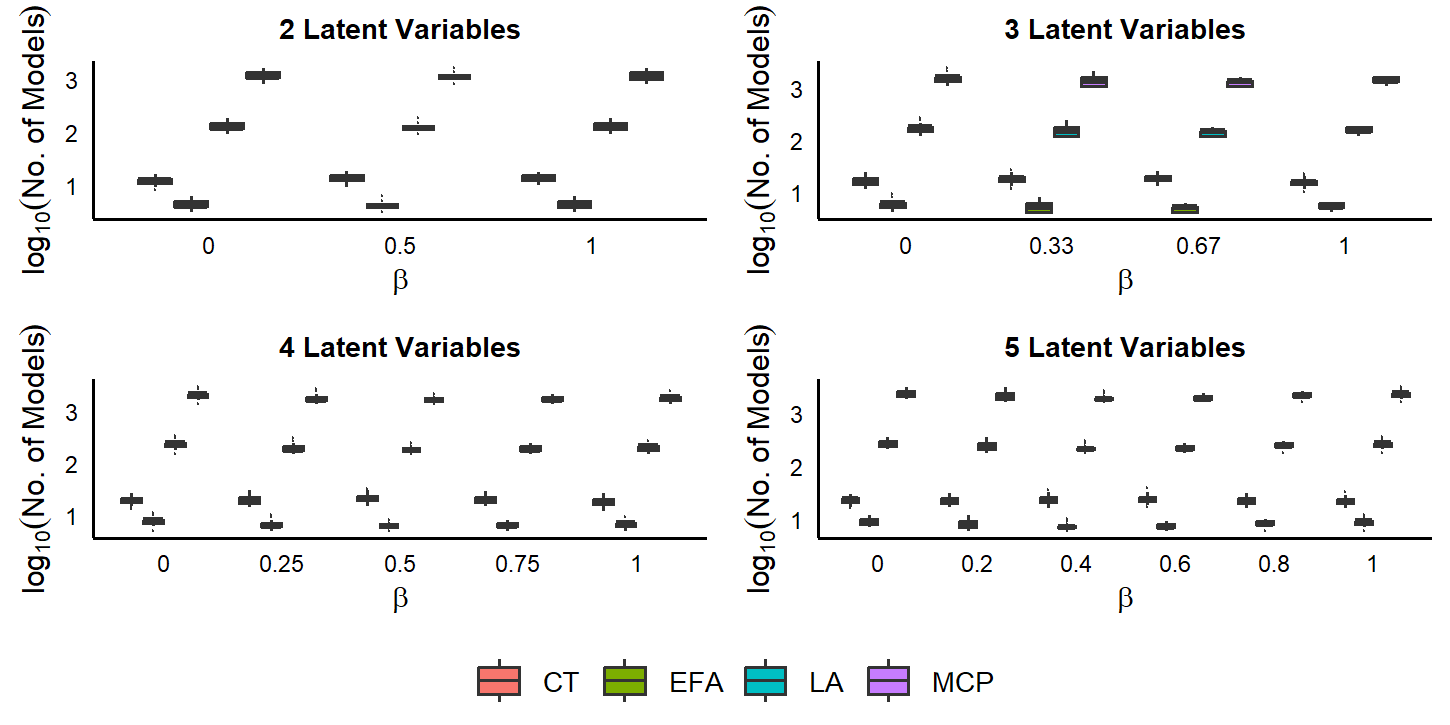}
  \caption{Boxplots of computational efficiency statistics for the low-dimensional simulation when $n = 1000$, as a function of $\beta$.}
\end{figure}

\begin{figure}[h!]
  \centering
  \includegraphics[width=.9\textwidth, keepaspectratio]{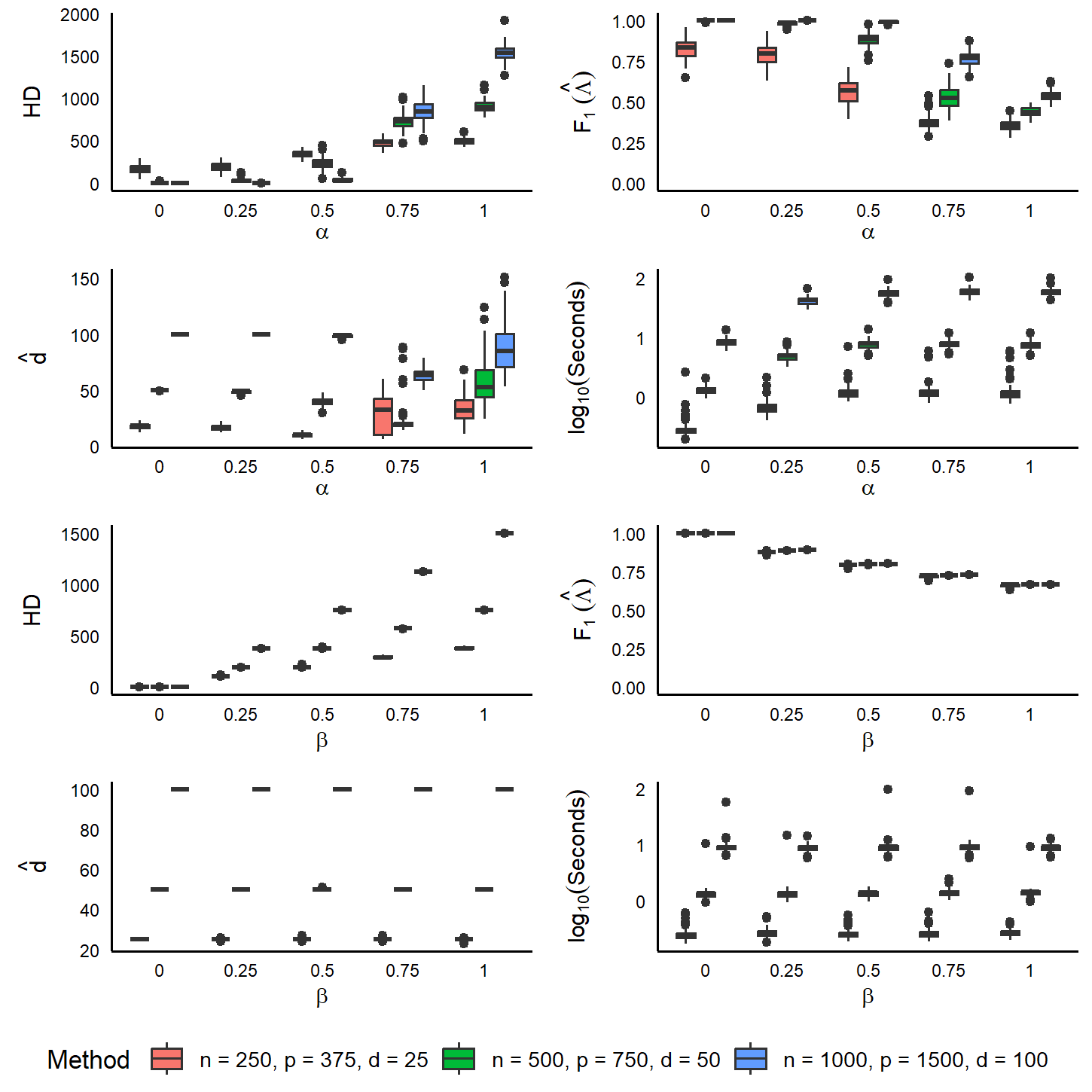}
  \caption{Boxplots of structural accuracy and computational efficiency statistics for the high-dimensional simulation when $n = 1000$, as a function of $\alpha$ and $\beta$.}
\end{figure}

\end{document}